\documentclass[lettersize,journal]{IEEEtran}
\usepackage{amsmath,amsfonts}
\usepackage{algorithmic}
\usepackage{tcolorbox}
\tcbuselibrary{skins, breakable}

\newtcolorbox[auto counter]{finding}[1][]{%
    colback=gray!10,
    colframe=gray!80,
    leftrule=3pt,
    rightrule=0pt,
    toprule=0pt,
    bottomrule=0pt,
    arc=0pt,
    boxsep=0pt,
    left=8pt,
    right=8pt,
    top=8pt,
    bottom=8pt,
    title={\textbf{Finding \thetcbcounter}:},
    fonttitle=\bfseries,
    coltitle=black,
    attach title to upper={\ },
    #1
}

\usepackage{amssymb}
\usepackage{pifont}

\usepackage{amssymb}
\usepackage{pifont}
\usepackage{algorithm}
\usepackage{array}
\usepackage[caption=false,font=normalsize,labelfont=sf,textfont=sf]{subfig}
\usepackage{textcomp}
\usepackage{stfloats}
\usepackage{url}
\usepackage{verbatim}
\usepackage{graphicx}
\usepackage{cite}
\usepackage{tabularx}
\usepackage{makecell}
\usepackage{booktabs}
\usepackage{multirow}
\usepackage{threeparttable}
\usepackage{url}
\usepackage{pdflscape}   
\usepackage{booktabs}    
\usepackage{multirow}
\usepackage{caption}     
\usepackage{mdframed}
\usepackage{hyperref}
\hypersetup{
  colorlinks=true,
  linkcolor=black,
  urlcolor=black,
  citecolor=black,
  bookmarksnumbered,
  unicode
}

\usepackage{graphicx} 
\usepackage{tcolorbox}
\tcbuselibrary{skins, breakable}

\usepackage[table]{xcolor}
\usepackage{array}
\usepackage{adjustbox}
\usepackage{pifont}
\usepackage{enumerate}
\usepackage{enumitem}
\usepackage{expdlist}

\begin{document}

\title{An Extensive Empirical Study on Code Translation Techniques}

\author{Ruihang Fan, Jiajun Jiang, Xinpeng Wang, Jiateng Fu, Fengjie Li and Jiasi Shen
\thanks{Jiajun Jiang is the corresponding author.}
\thanks{Ruihang Fan, Jiajun Jiang, Xinpeng Wang, Jiateng Fu and Fengjie Li are with School of Computer Software, College of Intelligence and Computing, Tianjin University, Tianjin 300350, China. Jiasi Shen is with the Department of Computer Science and Engineering, The Hong Kong University of Science and Technology, Hong Kong 999077, China. 

(E-mail: \{frexvite, jiangjiajun, wxp3023244137, world\_2004, fengjie\}@tju.edu.cn; sjs@cse.ust.hk).}}


\markboth{Journal of \LaTeX\ Class Files,~Vol.~14, No.~8, August~2021}%
{Shell \MakeLowercase{\textit{et al.}}: A Sample Article Using IEEEtran.cls for IEEE Journals}


\maketitle

\newcommand{\jiajun}[1]{\textcolor{cyan}{[Jiajun: #1]}}

\newcommand{\fengjie}[1]{\textcolor{orange}{[Fengjie: #1]}}

\newcommand{\jiasi}[1]{\textcolor{green}{[Jiasi: #1]}}

\newcommand{\artifact}{\url{https://doi.org/10.5281/zenodo.21973888}}

\newcommand{\VIMPT}[1]{VIM-PT}
\newcommand{\StructCoder}[1]{StructCoder}
\newcommand{\TranscoderST}[1]{TransCoder-ST}
\newcommand{\IT}[1]{InterTrans}
\newcommand{\ITMagi}[1]{IT(Magi)}
\newcommand{\ITStar}[1]{IT(Star)}
\newcommand{\UniTrans}[1]{UniTrans}
\newcommand{\ExeCoder}[1]{ExeCoder}

\newcommand{\distance}{3pt}
\setlength{\textfloatsep}{\distance}
\setlength{\floatsep}{\distance}
\setlength{\dbltextfloatsep}{\distance} 
\setlength{\dblfloatsep}{\distance} 
\begin{abstract}
Automated code translation is increasingly important for software evolution, yet the relative strengths and limitations of learning-based and large language model (LLM)-based techniques remain insufficiently understood. To address this gap, we conduct a large-scale empirical study comparing representative code translation techniques across methodological paradigms and translation granularities. We evaluate learning-based methods, LLM-based methods, and general-purpose LLMs on multilingual method-level and class-level benchmarks involving multiple programming languages. Our analysis considers executable correctness, code similarity, translation direction, translation granularity, and failure patterns. The results show that LLMs and LLM-based methods generally outperform learning-based methods in method-level correctness, although similarity metrics alone do not reliably reflect functional correctness. Translation direction substantially affects performance, particularly when translating between languages with different type-system characteristics. Class-level translation remains considerably more difficult than method-level translation because it requires preserving global semantics, interfaces, member relationships, and cross-method dependencies. Our error analysis further shows that static semantic errors and logical errors are the primary challenges in existing code translation systems. These findings provide empirical evidence and practical guidance for developing more robust, type-aware, structure-aware, and context-aware code translation techniques.
\end{abstract}


\maketitle

\section{Introduction}
\label{sec:intro}
\IEEEPARstart{C}{ode}
translation, the process of automatically converting source code into functionally equivalent versions in other programming languages, has become a cornerstone of modern software evolution. It underpins critical tasks such as cross-platform program migration~\cite{sdatrans,mppsmt,aura}, legacy system modernization~\cite{miningapi,vert}, and library upgrades~\cite{alphatrans,rustrepotrans}. As software systems evolve rapidly, the demand for such automated translation has grown substantially, since manual translation remains resource-intensive and time-consuming, while automation techniques can significantly reduce developer effort and maintenance costs~\cite{nguyen,transcoder}. Consequently, automated techniques have emerged to address this need.

Over the past few decades, a plethora of approaches have been proposed, ranging from early rigid, rule-based methods grounded in abstract syntax trees and pattern matching~\cite{sharpen,c2rust,duoglot,j2csharp,j2p,cxgo}, to sophisticated deep learning models~\cite{codebert,codet5,codet5+,structcoder,vimpt,transcoder-s,natgen,transcoder-st,contrabert}, and, most recently, techniques based on large language models (LLMs)~\cite{intertrans,cotran,unitrans,transagent}. In recent years, methods from different paradigms have achieved impressive results according to their respective metrics, yet the rapid pace of development has outpaced our understanding of their relative strengths and weaknesses. Although several empirical studies on code translation techniques have been conducted~\cite{Chen2025ASL, classeval, gtrans-eval, litrans, e2r, polyhumaneval}, they have typically adopted tools from only a single paradigm, i.e., either exclusively LLM-based methods or purely neural translation approaches. Consequently, there remains a lack of systematic, large-scale empirical research that performs genuine cross-paradigm, cross-task level, and cross-dataset comparisons of existing code translation techniques under unified settings, in order to assess the current state of the art.

To fill this gap, we conduct an extensive empirical study to systematically evaluate representative advanced techniques. Specifically, we first survey existing automated code translation approaches, primarily published between 2020 and 2025, and apply rigorous selection criteria, ultimately identifying \textbf{11} state-of-the-art representative techniques. These include four deep learning based methods (\VIMPT{}, \StructCoder{}, \TranscoderST{}, and a general-purpose code model, CodeT5+), four LLM-based methods (two variants of \IT{}, \UniTrans{}, and \ExeCoder{}), along with three recent general-purpose LLMs (GPT-3.5, GPT-4o-mini, and DeepSeek-V4-Flash). Finally, we conduct experiments on two method-level multilingual benchmarks and one class-level translation benchmark (G-TransEval, TransCoder-Uni, and ClassEval-T) to enable comprehensive and fair comparisons across code translation tasks involving five programming languages (C++, C\#, Java, JavaScript, and Python). Through in-depth analysis of the experimental results, we derive a set of findings, among which we highlight the following findings:

\begin{itemize}[leftmargin=*]
    \item \textbf{LLM-based methods generally achieve better correctness on method-level translation.} LLMs and LLM-based methods generally outperform learning-based methods in CA and CSR. However, improvements over the corresponding base LLM are not uniform and depend on the method design, the capability of the base model, and the target task. The advantage of LLM-based methods is also less pronounced when measured by CodeBLEU.

    \item \textbf{Translation direction affects correctness.} The direction of code translation has a clear impact on translation performance. Translating from statically typed languages to dynamically typed languages generally achieves better results than translating in the opposite direction, whereas the reverse direction tends to perform worse. This effect is also influenced to some extent by the inherent characteristics of the benchmark dataset.

    \item \textbf{Class-level translation remains substantially more difficult than isolated method-level translation.} On ClassEval-T, methods obtain relatively high CSR but much lower CA, indicating that producing compilable class-level code is considerably easier than preserving class semantics, member relationships, and behavioral equivalence.

    \item \textbf{Language effects are pronounced in class-level translation.} Within the class-level setting, translations targeting Python generally achieve higher CA than translations targeting Java or C++, while translation into C++ is particularly difficult in several directions. Larger contexts and cross-member dependencies amplify the challenges introduced by language-specific type and structural constraints.

    \item \textbf{Static semantic and logical errors are the dominant failure categories.} Among the sampled method-level failed cases, static semantic errors account for 49.37\% and logical errors account for 27.46\%, together comprising 76.83\% of all failures. At the class level, static semantic errors account for 70.67\% of the sampled failures. Function signature violations are the most common static semantic error, highlighting the difficulty of type mapping, interface preservation, symbol binding, and logical equivalence.
\end{itemize}

These findings lead to several implications for future research. In particular, code translation systems should move beyond isolated method-level translation, combine correctness-oriented and similarity-oriented evaluation, explicitly address direction-specific type inference and type mapping, and strengthen semantic verification for larger translation units.

In summary, this paper makes the following contributions:

\begin{itemize}[leftmargin=*]
    \item To the best of our knowledge, we conduct the \textbf{first large-scale empirical study} that evaluates code translation techniques across multiple methodological paradigms and translation granularities, including \textbf{11} representative techniques spanning learning-based methods, LLM-based methods, and general-purpose LLMs under unified experimental settings.

    \item We perform a \textbf{cross-paradigm, cross-dataset, and cross-granularity evaluation} using two method-level multilingual benchmarks and one class-level translation benchmark. This enables controlled comparisons across translation methods, programming languages, translation directions, and translation units.

    \item We provide a \textbf{multi-dimensional analysis of code translation performance} based on correctness and similarity metrics, including an investigation of language-direction effects, benchmark characteristics, and the differences between method-level and class-level translation.

    \item We conduct a \textbf{systematic manual analysis of translation failures}. We annotate 570 method-level failed outputs and 150 class-level failed outputs using a structured error taxonomy, with high inter-rater agreement measured by Fleiss' kappa. The analysis identifies static semantic errors, logical errors as major challenges.

    \item We summarize a \textbf{set of findings and implications} for developing more robust, type-aware, and context-aware code translation techniques, and open-source the experimental results and associated implementations to facilitate replication and future comparison.
\end{itemize}

\section{Related Work}

\subsection{Code Translation Methods}
\label{sec:relatedmethod}
Numerous code translation methods have been proposed to automatically convert source code written in one programming language into another functionally equivalent code in another language. Existing approaches can be broadly categorized into \textit{rule-based}, \textit{learning-based}, and \textit{LLM-based} methods. We also provide a brief summary of the recent representative methods in Table~\ref{tab:candidate}, showing the publishing year, supported languages, the benchmarks used for evaluation, and whether the source code is open-source.

\textbf{Rule-based methods} perform translation through hand-crafted rules that explicitly map source-language constructs to their counterparts in the target-language. Represent works include MetaLift~\cite{bhatia2023building}, DuoGlot~\cite{duoglot}, CxGo~\cite{cxgo}, Sharpen~\cite{sharpen} and C2Rust~\cite{c2rust}.
These methods typically operate by parsing source code into structured intermediate forms and then applying predefined transformation rules to produce target-language code. However, most of them are language-specific and require substantial manual effort to construct and maintain the translation rules. As a result, they are difficult to adapt to newly introduced features, evolving language standards, and the translated code may still require manual refinement~\cite{Verhoef}. 

\textbf{Learning-based methods} learn translation patterns from data instead of relying on hand-crafted rules. Early machine learning–based approaches to code translation were largely inspired by statistical machine translation (SMT).
Nguyen et al.~\cite{nguyen} investigated whether SMT techniques developed for natural languages could be adapted to programming languages.
Karaivanov et al.~\cite{kara} explored phrase-based methods and demonstrated the feasibility of SMT-based code translation.
Nguyen et al.~\cite{mppsmt} further proposed mppSMT, a divide-and-conquer method that achieved strong performance at that time.

Later, deep neural networks were introduced for code translation. 
Chen et al.~\cite{tree2tree} proposed a tree-based neural model for translating between JavaScript and CoffeeScript.
However, such supervised approaches are often constrained by the availability and quality of parallel corpora. 
To address this limitation, Roziere et al.~\cite{transcoder} introduced TransCoder, a fully unsupervised neural transcompiler. 
And follow-up works further improved it with unit test filtering and intermediate representations~\cite{transcoder-st,transcoder-ir}. 

\newcommand{\cmark}{\ding{51}} 
\newcommand{\xmark}{\ding{55}} 
\newcolumntype{C}[1]{>{\centering\arraybackslash}m{#1}}

\newcolumntype{P}[1]{>{\raggedright\arraybackslash}m{#1}}

\begin{table}[tbp]
  \centering
  \caption{A brief summary of code translation techniques.}
  \label{tab:candidate}
  \resizebox{\columnwidth}{!}{
  \footnotesize
  \begin{threeparttable}
  \begin{tabular}{c|P{2.3cm}C{0.5cm}C{1.9cm}C{1.5cm}C{0.6cm}}
    \toprule
     & \textbf{Name} & \textbf{Year$\uparrow$} & \textbf{Supported Languages} & \textbf{Benchmark} & \textbf{OSS} \\
    \midrule
    \multirow{17}{*}{\centering \rotatebox{90}{Learning-based}}
    & \text{TransCoder}~\cite{transcoder} & 2020 & C++, Java, Python & D1 & \cmark \\
    \cline{2-6}
    & \vspace{1pt}PLBART~\cite{plbart} & 2021 & C\#, Java & D4 & \cmark \\
     \cline{2-6}
    & \TranscoderST{}~\cite{transcoder-st}$^\bigstar$ & 2021 & C++, Java, Python & D1 & \cmark \\
    \cline{2-6}
    & \text{TransCoder-IR}~\cite{transcoder-ir} & 2022 & C++, Go, Java, Rust & D1 & \cmark \\
    \cline{2-6}
    & \vspace{1pt}NatGen~\cite{natgen} & 2022 & C\#, Java & D4 & \cmark \\
    \cline{2-6}
    & \vspace{1pt}Contrabert~\cite{contrabert} & 2023 & C\#, Java & D4 & \cmark \\
    \cline{2-6}
    & SDA-Trans~\cite{sdatrans} & 2023 & C++, Java, Python & D1 & \xmark \\
    \cline{2-6}
    & SumGenToBT~\cite{sumgentobt} & 2023 & Java, Python & D1 & \cmark \\
    \cline{2-6}
    & \StructCoder{}~\cite{structcoder}$^\bigstar$ & 2024 & C\#, Java & D4 & \cmark \\
    \cline{2-6}
    & \VIMPT{}~\cite{vimpt}$^\bigstar$ & 2024 & {C/C++, C\#, Java, JavaScript, Python, PHP} & D5 & \cmark \\
    
    \midrule
    \multirow{14}{*}{\centering \rotatebox{90}{LLM-based}}
    & CodeT5~\cite{codet5} & 2021 & multiple & D4 & \cmark \\
    \cline{2-6}
    & CodeT5+~\cite{codet5+}$^\bigstar$ & 2023 & Multiple & D4 & \cmark \\
    \cline{2-6}
    & CodeGeeX~\cite{codegeex} & 2023 & Multiple & D8, D6 & \cmark \\
    \cline{2-6}
    & \vspace{1pt}FSCTrans~\cite{fsctrans} & 2024 & C\#, Java, Python & D1, D4 & \cmark \\
    \cline{2-6}
    & \vspace{1pt}Mftcoder~\cite{mftcoder} & 2024 & Multiple & CodeFuseEval & \cmark \\
    
    \cline{2-6}
    & \vspace{1pt}CodeFuse-13B~\cite{codefuse} & 2024 & Multiple & CodeFuseEval & \cmark \\
    \cline{2-6}
    & TransAGENT~\cite{transagent} & 2024 & Multiple & TransAGENT-test & \xmark \\
    \cline{2-6}
    & \vspace{1pt}CoTran~\cite{cotran} & 2024 & Multiple & D10 & \cmark \\
    \cline{2-6}
    & \UniTrans{}~\cite{unitrans}$^\bigstar$ & 2024 & Multiple & D9 & \cmark \\ 
    \cline{2-6}
    &  \vspace{1pt}\IT{}~\cite{intertrans}$^\bigstar$ & 2025 & Multiple & D8, D1, D2 & \cmark \\ 
     \cline{2-6}
    & \vspace{1pt}ExeCoder~\cite{execoder}$^\bigstar$ & 2025 & C++, Java, Python & TransCoder-test-X & \cmark \\
    
    \midrule
    \multirow{7}{*}{\centering \rotatebox{90}{\makecell{Rule-based}}} 
    & Sharpen~\cite{sharpen} & 2008 &  C\#, Java & - & \cmark \\
    \cline{2-6}
    &  \vspace{1pt}C2Rust~\cite{c2rust} & 2018 & C++, Rust & - & \cmark \\
    \cline{2-6}
    & \vspace{1pt}CxGo~\cite{cxgo} & 2021 & C++, Go & - & \cmark \\
    \cline{2-6}
    & DuoGlot~\cite{duoglot} & 2023 & Python, JavaScript & D1 & \cmark \\
    \cline{2-6}
    & \vspace{1pt}MetaLift~\cite{bhatia2023building} & 2023 & C++, Java & Spark, Domino & \cmark \\
     \bottomrule
  \end{tabular}
  \begin{tablenotes}
    \small
    \item ``$\bigstar$'': studied methods in this paper. ``-'': the tool did not explicitly evaluate on any benchmarks, and ``Dk'' denotes dataset shown in Table~\ref{tab:candidate_datasets}.
  \end{tablenotes}
  \end{threeparttable}
  }
\end{table}


\textbf{LLM-based methods} have recently emerged as a promising direction for code translation. Unlike earlier learning-based approaches that mainly depend on task-specific training, these methods typically leverage the reasoning and iterative refinement abilities of LLMs to improve translation quality. 
For example, 
Jana et al.~\cite{cotran} presented CoTran, which fine-tunes an LLM with reinforcement learning guided by compiler and symbolic execution feedback, directly optimizing for compilation correctness and functional equivalence. 
Zheng et al.~\cite{codegeex} introduced CodeGeeX, a large multilingual pre-trained model capable of both code generation and translation across 23 programming languages, with its evaluation benchmark HumanEval-X extending beyond Python to multiple languages.
Yang et al.~\cite{unitrans} proposed UniTrans, which enhances translation through automated test-driven feedback. 
Yuan et al.~\cite{transagent} introduced TranAgent, a multi-agent framework that uses multiple LLMs to correct syntactic and semantic errors during translation. 
Macedo et al.~\cite{intertrans} developed InterTrans, which leverages intermediate languages to bridge syntactic gaps between source and target languages. 
Additionally, a wide range of LLMs for code generation can also be applied to code translation tasks, such as CodeBERT~\cite{codebert},  CodeT5~\cite{codet5}, CodeT5+~\cite{codet5+}, and MftCoder~\cite{mftcoder}. Several of these models have achieved strong performance on code translation benchmarks and are even regarded as competitive baselines—CodeT5+ being a notable example.

\newcolumntype{C}[1]{>{\centering\arraybackslash}m{#1}}
\newcolumntype{L}[1]{>{\raggedright\arraybackslash}m{#1}}

\begin{table}[htbp]
  \centering
  \caption{Benchmarks for code translation evaluation.}
  \label{tab:candidate_datasets}
  
  \resizebox{\columnwidth}{!}{
  \begin{threeparttable}
  \begin{tabular}{L{3.4cm}cC{3.2cm}cc}
    \toprule
    \textbf{Name} & \textbf{Year$\uparrow$} & \textbf{Languages} & \textbf{Probs.} & \textbf{Tests} \\
    \midrule
    D1. TransCoder-test~\cite{transcoder} & 2020 & C++, Java, Python & 948  & \cmark \\
    \midrule
    D2. CodeNet~\cite{codenet} & 2021 & C/C++, Go, Java, Python & 200 & \xmark \\
    \midrule
    D3. AVATAR~\cite{avatar} & 2021 & Java, Python & 250 & \xmark \\
    \midrule
    D4. CodeXGLUE~\cite{codexglue} & 2021 & C\#, Java & 1000 & \xmark \\
    \midrule
    D5. CoST~\cite{Cost} & 2022 & C/C++, C\#, Java, JavaScript, Python, PHP & 1625  & \xmark \\
    \midrule
    D6. XLCoST~\cite{xlcost} & 2022 & C/C++, C\#, Java, JavaScript, Python, PHP & 11265 & \cmark \\
    \midrule
    D7. \textbf{G-TransEval}~\cite{gtrans-eval} & 2023 & C++, C\#, Java, JavaScript, Python & 400 & \cmark \\
    \midrule
    D8. HumanEval-X~\cite{codegeex} & 2023 & C++, Go, Java, JavaScript, Python & 164  & \cmark \\
    \midrule
    D9. \textbf{TransCoder-Uni}~\cite{unitrans} & 2024 &  C++, Java, Python & 568 & \cmark \\
     \midrule
    D10. AVATAR-TC~\cite{avatar-tc} & 2024 & Java, Python & 1746 & \cmark \\
    \midrule
    D11. \textbf{ClassEval-T}~\cite{classeval} & 2025 & C++, Java, Python& 94 & \cmark \\
   
    
    \bottomrule
  \end{tabular}
  \begin{tablenotes}
    \small
    \item *``Probs'' denotes the number of programming problems covered in the dataset.
  \end{tablenotes}
  \end{threeparttable} 
  }
\end{table}

\subsection{Code Translation Benchmarks}
\label{sec:relatebench}

Over the past few years, several benchmarks have been proposed for evaluating code translation methods. 
CodeNet~\cite{codenet} provides large-scale parallel programs collected from online coding platforms, covering 55 different programming languages.
Ahmad et al. introduce Avatar~\cite{avatar}, which focuses on more challenging Python--Java translation tasks.
CodeXGLUE~\cite{codexglue} provides a code-to-code translation benchmark based on parallel Java--C\# programs.
Although they cover a wide range of programming languages, many of them lack executable test cases, which limits the reliable evaluation of the functional correctness of translated code. 

More recent benchmarks address this issue by providing executable test suites.
Roziere et al.~\cite{transcoder-st} proposed a benchmark (referred to as TransCoder-Uni) from GeeksforGeeks programs in Python, C++, and Java, and 568 out of 948 cases include test suites.
Jiao et al.~\cite{gtrans-eval} introduced G-TransEval, which contains 400 parallel translation instances across five languages, along with difficulty labels and complete test suites for every language pair.
Xue et al.~\cite{classeval} proposed ClassEval-T, a class-level code translation benchmark covering Python, C++, and Java, featuring longer code snippets, complex dependencies, and comprehensive test suites. We list several recognized benchmarks in the field of code translation in Table~\ref{tab:candidate_datasets}. After our verification, all benchmarks in the table have been used by at least two works in the code translation domain. 

\subsection{Empirical Studies on Code Translation}



To investigate the effectiveness of existing code translation techniques, several studies have been conducted. 
Jiao et al.~\cite{gtrans-eval} proposed a four-type taxonomy for code translation and systematically evaluated state-of-the-art neural translation models across these translation types. 
Gong et al.~\cite{gong2025tracy} studied the execution efficiency of LLM-based code translation,
and Pan et al.~\cite{litrans} studied the effectiveness of LLMs for automated code translation and analyzed LLM-introduced translation bugs. 
\UniTrans{}~\cite{unitrans} further enhanced code translations through test generation, execution-based checking, and iterative repair.
Tao et al.~\cite{polyhumaneval} further introduced PolyHumanEval and used it to conduct a large-scale empirical study of LLM-based code translation.
However, these studies either focus on neural or LLM-based (\textbf{especially focus on end-to-end LLM methods}) methods or are limited to a small set of programming languages and evaluation perspectives. In contrast, our work aims to compare both recent learning-based and LLM-based methods under a unified setting and provides a more comprehensive evaluation from multiple perspectives, including quantitative metrics and error type analysis. 
\section{Methodology}

\subsection{Code Translation Method Selection}
\label{sec:studiedmethod}

As mentioned in the introduction, existing studies predominantly focus on individual paradigms (e.g., LLM-based methods, especially focus on end-to-end LLMs), while a systematic comparison of different approaches under a unified setting remains underexplored. As discussed in Section~\ref{sec:relatedmethod}, numerous code translation methods have been proposed in the past decade. This study aims to comprehensively compare the performance of the latest state-of-the-art techniques. Accordingly, we focus on methods published between 2020 and 2025. We further limit our scope to open-source approaches, primarily for two reasons: (1) reproducing a large number of methods would require significant time and effort, and (2) reimplementations often fail to achieve the performance reported in the original papers, potentially confounding the evaluation of each method’s intrinsic capabilities. Finally, we selected the representative methods from two categories, i.e., learning-based and LLM-based approaches (listed in Table~\ref{tab:candidate}), to ensure the feasibility of our study and the reliability of our experimental results. To identify the most representative approaches within these categories, we applied the following three selection criteria.


\begin{enumerate}[leftmargin=*]
    \item \textbf{Performance (outstanding performance)}: Among methods developed under comparable timelines, we prioritize those achieving best performance on at least one benchmark, ensuring that selected systems genuinely represent the current best performance within their respective technical paradigms.
    \item \textbf{Recency (latest techniques)}: Higher preference is given to models and methods released or updated after 2024, so that our experiments can reflect the latest advancements in the field of code translation.
    \item \textbf{Language diversity (applicable to diverse languages)}: 
    We favor methods supporting the largest number of programming languages, aiming to obtain more comprehensive and generalizable experimental results.
\end{enumerate}

Based on the above criteria, we ultimately selected \textbf{eight} state-of-the-art code translation methods from existing studies. These methods include four learning-based approaches (three translation-specific methods: \VIMPT{}, \StructCoder{}, \TranscoderST{}, and one general-purpose code model: CodeT5+), as well as four LLM-based methods (two variants of \IT{}, \UniTrans{}, \ExeCoder{}). The rationale is as follows: VIM-PT, StructCoder, UniTrans, ExeCoder and InterTrans are the best-performing open-source methods in recent years in the fields of learning-based code translation and LLM-based code translation, respectively (TransAGENT surpasses UniTrans but is not open-source). TransCoder-ST and CodeT5+ have been employed as classic competitive baselines in prior code translation studies, and we include them in our experiments to obtain more representative results~\cite{fsctrans,structcoder,unitrans,cotran}. In addition, to conduct a comprehensive evaluation, we also incorporate \textbf{three} recent general-purpose LLMs into our study, which have demonstrated outstanding performance across various software engineering tasks~\cite{aprllm,giantrepair,aprpllm}, including two commercial LLMs (GPT-3.5-turbo and GPT-4o-mini) and one open-source LLM (DeepSeek-V4-Flash). Consequently, we investigate \textbf{11} representative approaches in our evaluation, which (except for LLMs) are marked with $\bigstar$ in Table~\ref{tab:candidate} and described in detail below.

\begin{description}[\setleftmargin{10pt}]
    \item[\VIMPT{}~\cite{vimpt}:] A unified multilingual program translation model based on variational inference and information disentanglement. It jointly learns language-shared semantics and language-specific features, effectively leveraging non-parallel and partially missing data to significantly improve accuracy and generalization in cross-language code translation.
    \item[\StructCoder{}~\cite{structcoder}:] A structure-aware Transformer model that explicitly incorporates abstract syntax trees (ASTs) and data flow graphs (DFGs) into both encoder and decoder. It further enhances performance on code translation and text-to-code generation by introducing two auxiliary tasks in the decoder: AST path prediction and data flow prediction.
    \item[\TranscoderST{}~\cite{transcoder-st}:] An unsupervised code translation approach based on self-training. It automatically generates high-coverage unit tests for source code and uses these tests to filter semantically equivalent translations, thereby constructing high-quality pseudo-parallel corpora for iterative model refinement.

    \item[CodeT5+~\cite{codet5+}:] A model featuring that supports encoder-only, decoder-only, and encoder-decoder modes within a unified architecture. It is trained with multi-stage objectives, including span denoising, causal language modeling, contrastive learning, and text-code matching, and achieves strong performance across code understanding and generation tasks.
    
    \item[\IT{}~\cite{intertrans}:] An LLM-based code translation method that constructs an ``intermediate-language translation path'' and employs tree search combined with test verification to substantially improve functional correctness in cross-language translation. In our experiments, we use InterTrans implementations based on Magicoder-SDS-7B~\cite{magicoder} and StarCoder2-15B~\cite{starcoder}, referred to hereafter as \textbf{\ITMagi{}} and \textbf{\ITStar{}}, respectively.
    
    \item[\UniTrans{}~\cite{unitrans}:] A general-purpose LLM-based code translation framework that validates and repairs translation outputs by generating and executing test cases. It further iteratively refines prompts using error feedback to enhance both functional correctness and robustness in cross-language code translation.

    \item[\ExeCoder{}~\cite{execoder}:] An LLM specifically designed for code translation, aimed at utilizing executability representations such as functional semantics, syntax structures, and variable dependencies to enhance the capabilities of LLMs in code translation. 

    \item[GPT-3.5-turbo~\cite{gpt35turbo}:] An efficient and cost-balanced LLM from OpenAI. It was widely used as a baseline for code understanding and generation tasks in previous studies~\cite{unitrans}. Specifically, we adopted the version of GPT-3.5-turbo-1106 in our experiment.
    
     \item[GPT-4o-mini~\cite{gpt-4o-mini}:] A lightweight OpenAI model offering low latency and strong reasoning capabilities, ideal for real-time applications and code generation. Specifically, we employed the version of GPT-4o-mini-2024-07-18 in our experiment.
    
    \item[DeepSeek-V4-Flash~\cite{deepseek-v4-f}:] A high-performance model from DeepSeek featuring a Mixture-of-Experts architecture, delivering strong results in code and reasoning tasks with high cost-effectiveness.
    
\end{description}



\subsection{Benchmark Selection}

As introduced in Section~\ref{sec:relatebench}, a number of benchmarks for code translation have been proposed, as shown in Table~\ref{tab:candidate_datasets}. Similarly, we also applied several selection criteria analogous to those used for the studied method selection:

\begin{enumerate}[leftmargin=*]
    \item \textbf{Multilingual (covering diverse languages)}: To enable a comprehensive comparison among diverse methods and investigate their translation performance when applied to different programming languages of the same task (A task denotes a distinct function regardless of implementation languages), we prefer the benchmarks that have more tasks and
    involve more parallel implementations of different languages.
    \item \textbf{Verifiability (including test suite)}: Since the same functionality can be implemented in different ways, directly comparing the equivalence of the translated code to the ground-truth implementation is insufficient. To enable a more in-depth analysis and make the evaluation results verifiable in an automatic manner, we require the benchmark including associated test suites, which play as the (incomplete) semantic specification of the target code, and thus can automatically and reliably examine the functional correctness of the translated code.
    \item \textbf{Compatibility (applicable to studied methods)}: We require that the benchmarks should accommodate as many of the selected methods introduced in Section~\ref{sec:studiedmethod}, facilitating more generalizable conclusions rather than findings limited to a single dataset.
\end{enumerate}


According to these selection criteria, we ultimately selected two representative method-level code translation benchmarks in this study, G-TransEval and TransCoder-Uni (the cleaned version of the TransCoder-ST dataset prepared by Yang et al.~\cite{unitrans}), as well as one class-level code translation benchmark, ClassEval-T~\cite{classeval}, involving 1,062 distinct translation tasks in total.
All three benchmarks feature comprehensive test suites and multilingual parallel implementations. For the method-level benchmarks, they are applicable to nearly all of our studied methods, while the class-level benchmark is also compatible with most large language models and LLM-based approaches (learning-based methods are difficult to adapt due to context length limits). We also emphasized these three datasets in bold in Table~\ref{tab:candidate_datasets}. We exclude XLCoST from test-time evaluation because it is used for fine-tuning the learning-based methods in our study (Section~\ref{sec:implementation}).

\section{Experimental Setup}
\subsection{Research Questions}
\label{sec:rq}
In this study, we aim to answer the following research questions:

\begin{description}[\setleftmargin{10pt}]
\item[RQ1:] \textbf{How do existing methods perform overall on method-level code translation tasks?} 
To answer this question, we conduct extensive experiments on two large-scale benchmarks and compare the overall correctness of all selected methods. Specifically, we evaluate the translated code using compilation and functional correctness metrics, and further analyze the results with CodeBLEU for a multi-dimensional view of translation quality.

\item[RQ2:] \textbf{How do existing methods perform on different translation tasks and language directions?}
Since the same method may behave differently when translating between different language pairs or in opposite directions, we further investigate task-level performance variations across source-target language combinations. This RQ aims to identify language-direction effects and examine how programming language characteristics influence translation performance.

\item[RQ3:] \textbf{How does code translation perform at the class level?}
Compared with method-level translation, class-level translation involves longer contexts, richer inter-method dependencies, and more complex type structures. In this RQ, we extend the evaluation to class-level code translation and study whether existing methods can still preserve correctness under more complex code contexts.

\item[RQ4:] \textbf{What kinds of translation errors do current methods commonly make?}
Although existing methods achieve strong performance overall, they still produce incorrect outputs. To better understand these failures, we manually analyze translation errors produced by the selected methods, categorize them into major error types, and examine their distributions across datasets and methods.
\end{description}

\subsection{Evaluation Metrics}
\label{sec:metrics}

We use three evaluation metrics: \textbf{Computational Accuracy (CA)}, \textbf{Compilation Success Rate (CSR)}, and \textbf{CodeBLEU (CB)}~\cite{codebleu}. Following prior work~\cite{transcoder,transcoder-st,unitrans}, CA measures the proportion of generated programs that pass all official test cases, while CSR measures the proportion that compile successfully. CodeBLEU measures the similarity between generated code and reference code by combining n-gram matching, keyword matching, abstract syntax tree matching, and data flow matching. We use the standard setting in which all four components are equally weighted.

We collectively refer to CA and CSR as \textbf{correctness metrics}, since they measure the functional and syntactic correctness of generated code. We refer to CodeBLEU as a \textbf{similarity metric}, since it evaluates how closely the generated code matches the reference implementation in textual and structural terms.
\begin{table}[tbp]
  \centering
  \resizebox{\columnwidth}{!}{
  \begin{threeparttable}
    \caption{Configurations of hyperparameters for model training.}
    \label{tab:hyperparameters}
    \begin{tabular}{lccc}
      \toprule
      \textbf{Hyperparameter} & \textbf{StructCoder} & \textbf{VIM-PT} & \textbf{CodeT5+} \\
      \midrule
      Parameter size      &  $ \sim $ 224M           &  $ \sim $ 222M      &  $ \sim $ 220M       \\
      Learning rate & 1e-5                 & 1e-4            & 5e-5             \\
      Batch size              & 25                   & 8               & 8                \\
      Gradient accumulation steps & /              & 2               & 4                \\
      Epoch                   & 25                   & 25\tnote{*}     & 10               \\
      \bottomrule
    \end{tabular}
    \begin{tablenotes}
      \small
      \item[*]VIM-PT includes 10 auto-encoding epochs and 15 formal epochs.
    \end{tablenotes}
  \end{threeparttable}
  }
\end{table}
\subsection{Implementation and Configuration}
\label{sec:implementation}
\textbf{Learning-based Methods.} For \TranscoderST{}~\cite{transcoder-st}, we used the best checkpoint provided by the original authors. 
For \VIMPT{}~\cite{vimpt}, \StructCoder{}~\cite{structcoder}, and CodeT5+~\cite{codet5+}, we followed prior work and trained/fine-tuned the models on the XLCoST~\cite{xlcost} dataset. In addition, because TransCoder-Uni and XLCoST are derived from the same data source, we removed from the training set any examples that overlap with our evaluation benchmarks. The training hyperparameters are listed in Table~\ref{tab:hyperparameters}, and all settings follow the default configurations of the corresponding open-source repositories.

For each method, we evaluated using the checkpoint that achieved the best performance on the validation set. This choice was motivated by two factors: 1) learning-based methods exhibit less randomness compared to LLMs; 2) repeated training would incur substantial computational costs (e.g., VIM-PT training took approximately 1,266 GPU hours).

\noindent\textbf{LLM-based Methods}. The officially released \UniTrans{}~\cite{unitrans} cannot be directly executed on G-TransEval~\cite{gtrans-eval} due to the tight coupling of part of its code with the source dataset. We therefore adapted the modules for general translation enhancement and translation repair to evaluate and rectify translation outputs generated by LLMs.

For each method, we follow prior works~\cite{classeval, unitrans} and set the hyperparameters to temperature = 0.8, top\_p = 0.9, and n = 1, i.e., only the first generated output was used. We then conducted three independent runs and reported the average results.


\noindent\textbf{Experimental environment} Our experiments were conducted on a local machine equipped with dual Intel Xeon 6388 CPUs, 512GB RAM, and two NVIDIA A800-SXM4-80GB GPUs, running Ubuntu 20.04.6LTS.

\section{Results Analysis}

\subsection{RQ1: Method-Level Code Translation Performance}

\begin{figure*}[tbp]
\centering
\includegraphics[width=2.0\columnwidth]{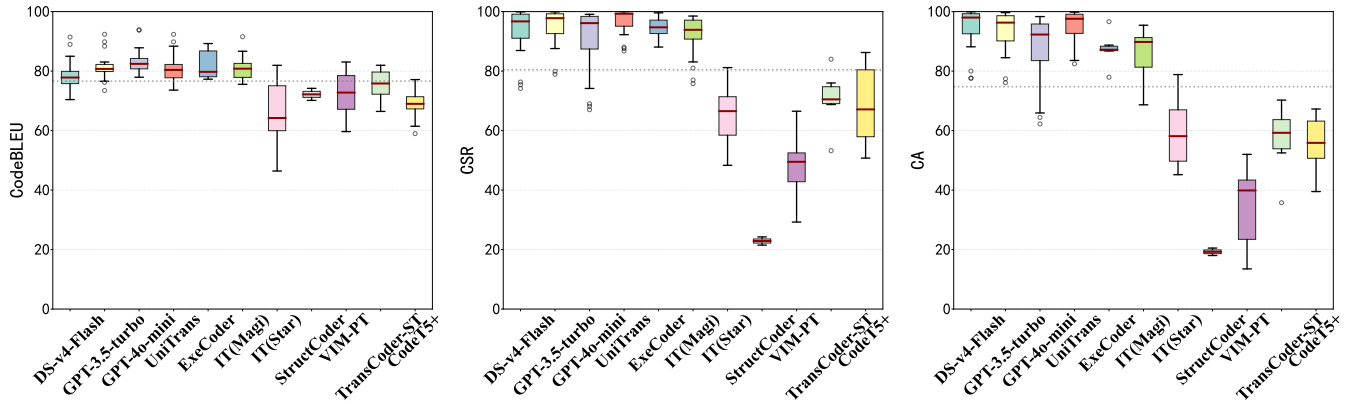}
\caption{Result distribution regarding all metrics for each method on G-TransEval.}
\label{figs:box1}
\end{figure*}

\begin{figure*}[tbp]
\centering
\includegraphics[width=2.0\columnwidth]{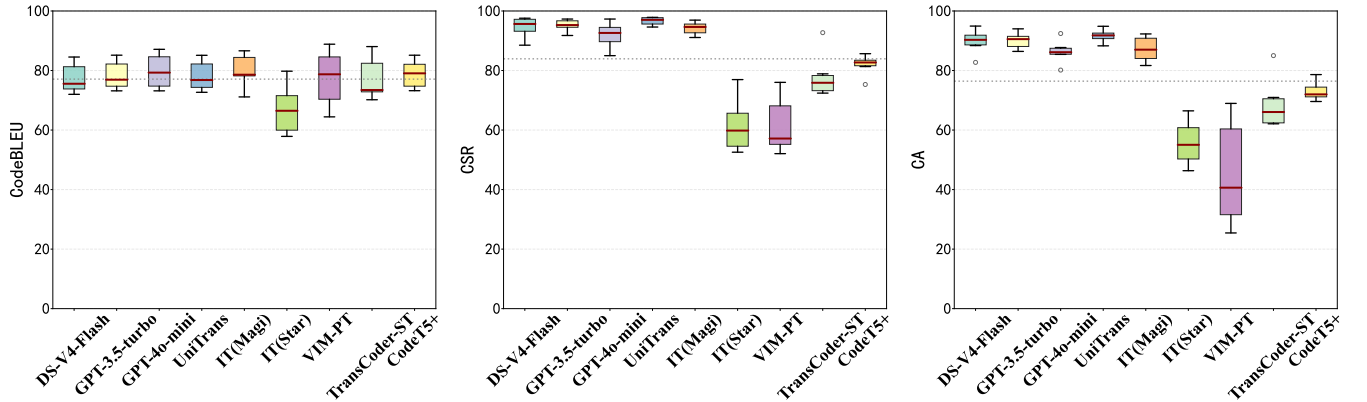}
\caption{Result distribution regarding all metrics for each method on TransCoder-Uni.}
\label{figs:box2}
\end{figure*}

In this research question, we evaluate the overall performance of different approaches on two large-scale benchmarks. Specifically, we consider the first output of each method by default, namely CA@1 and CSR@1, and further use CodeBLEU as a similarity metric for multi-dimensional evaluation. Table~\ref{tab:overall-gtrans} and Table~\ref{tab:overall-transuni} report the correctness results of different tools across all language pairs on G-TransEval~\cite{gtrans-eval} and TransCoder-Uni~\cite{unitrans}, respectively. Figure~\ref{figs:box1} and Figure~\ref{figs:box2} show the distributions of different metrics for all methods. Detailed CodeBLEU results are available on the project website.

\begin{table*}[htbp]
  \centering
  \scriptsize
  \begin{threeparttable}
  \caption{Translation performance on G-TransEval across methods. The table shows results for both translation directions (e.g., Java to C\# and C\# to Java) between each pair of languages (e.g., Java$\rightleftharpoons$C\#). Empty cells indicate that the corresponding method is not applicable. A color gradient is used, darker shades highlight larger values and thus better performance (\%).}
  \label{tab:overall-gtrans}
  \setlength{\tabcolsep}{2.5pt}
  \begin{tabular}{l *{20}{c} *{2}{c}}
    \toprule
    \multirow{2}{*}{Name} 
    & \multicolumn{2}{c}{Java$\rightleftharpoons$C\#} 
    & \multicolumn{2}{c}{Java$\rightleftharpoons$C++} 
    & \multicolumn{2}{c}{Java$\rightleftharpoons$Py.} 
    & \multicolumn{2}{c}{Java$\rightleftharpoons$JS} 
    & \multicolumn{2}{c}{C++$\rightleftharpoons$Py.}
    & \multicolumn{2}{c}{C++$\rightleftharpoons$JS} 
    & \multicolumn{2}{c}{C++$\rightleftharpoons$C\#} 
    & \multicolumn{2}{c}{Py.$\rightleftharpoons$C\#} 
    & \multicolumn{2}{c}{Py.$\rightleftharpoons$JS} 
    & \multicolumn{2}{c}{C\#$\rightleftharpoons$JS} 
    & \multicolumn{2}{c}{Average} \\
    \cmidrule(lr){2-3} \cmidrule(lr){4-5} \cmidrule(lr){6-7} \cmidrule(lr){8-9} \cmidrule(lr){10-11}
    \cmidrule(lr){12-13} \cmidrule(lr){14-15} \cmidrule(lr){16-17} \cmidrule(lr){18-19} \cmidrule(lr){20-21}
    \cmidrule(lr){22-23}
    & CSR & CA & CSR & CA & CSR & CA & CSR & CA & CSR & CA 
    & CSR & CA & CSR & CA & CSR & CA & CSR & CA & CSR & CA & CSR & CA \\
    \midrule

       & \cellcolor[HTML]{BFDBEE}{99.17} & \cellcolor[HTML]{FFCBCB}{98.92} & \cellcolor[HTML]{E8F0FA}{92.67} & \cellcolor[HTML]{FFE2E2}{92.67} & \cellcolor[HTML]{BEDBEE}{99.75} & \cellcolor[HTML]{FFCBCB}{99.58} & \cellcolor[HTML]{D0E4F2}{99.00} & \cellcolor[HTML]{FFCCCC}{97.42} & \cellcolor[HTML]{C0DCEF}{99.42} & \cellcolor[HTML]{FFCCCC}{99.25}
      & \cellcolor[HTML]{D4E6F1}{98.67} & \cellcolor[HTML]{FFCCCC}{96.92} & \cellcolor[HTML]{C9E0EF}{98.33} & \cellcolor[HTML]{FFCFCF}{98.08} & \cellcolor[HTML]{EDF4FB}{88.17} & \cellcolor[HTML]{FFE9E9}{86.92} & \cellcolor[HTML]{E6F0F9}{96.08} & \cellcolor[HTML]{FFD8D8}{93.33} & \cellcolor[HTML]{E0ECF6}{97.67} & \cellcolor[HTML]{FFD2D2}{96.50} & \multirow{2}{*}{94.06} & \multirow{2}{*}{92.86} \\
      \multirow{-2}{*}{DS-V4-Flash}
      & \cellcolor[HTML]{B8D8EC}{99.92} & \cellcolor[HTML]{FFCBCB}{99.92} & \cellcolor[HTML]{BEDBEE}{99.75} & \cellcolor[HTML]{FFCBCB}{99.75} & \cellcolor[HTML]{E8F0FA}{93.25} & \cellcolor[HTML]{FFE1E1}{91.75} & \cellcolor[HTML]{F0F5FC}{92.17} & \cellcolor[HTML]{FFE3E3}{88.92} & \cellcolor[HTML]{F8FBFE}{77.67} & \cellcolor[HTML]{FFF4F4}{76.33}
      & \cellcolor[HTML]{F9FBFE}{77.58} & \cellcolor[HTML]{FFF4F4}{74.17} & \cellcolor[HTML]{E8F0FA}{93.17} & \cellcolor[HTML]{FFE1E1}{92.83} & \cellcolor[HTML]{C4DEF0}{99.25} & \cellcolor[HTML]{FFCCCC}{99.08} & \cellcolor[HTML]{BEDBEE}{99.50} & \cellcolor[HTML]{FFCCCC}{99.50} & \cellcolor[HTML]{F8FBFE}{80.00} & \cellcolor[HTML]{FFF3F3}{75.42} \\

      & \cellcolor[HTML]{BFDBEE}{98.92} & \cellcolor[HTML]{FFCDCD}{98.58} & \cellcolor[HTML]{E4EFF8}{94.42} & \cellcolor[HTML]{FFE1E1}{94.25} & \cellcolor[HTML]{C0DCEF}{99.42} & \cellcolor[HTML]{FFCDCD}{98.92} & \cellcolor[HTML]{BEDBEE}{99.50} & \cellcolor[HTML]{FFCFCF}{97.50} & \cellcolor[HTML]{E0ECF6}{96.75} & \cellcolor[HTML]{FFD9D9}{96.17} 
      & \cellcolor[HTML]{BFDBEE}{99.25} & \cellcolor[HTML]{FFD5D5}{96.67} & \cellcolor[HTML]{C2DDEF}{98.83} & \cellcolor[HTML]{FFCDCD}{98.17} & \cellcolor[HTML]{E8F0FA}{89.42} & \cellcolor[HTML]{FFEAEA}{86.42} & \cellcolor[HTML]{E0ECF6}{96.08} & \cellcolor[HTML]{FFE2E2}{91.17} & \cellcolor[HTML]{C2DDEF}{98.83} & \cellcolor[HTML]{FFD6D6}{96.42} & \multirow{2}{*}{94.83} & \multirow{2}{*}{93.23} \\
      \multirow{-2}{*}{GPT-3.5-turbo} 
      & \cellcolor[HTML]{B8D8EC}{99.92} & \cellcolor[HTML]{FFCBCB}{99.67} & \cellcolor[HTML]{C0DCEF}{99.17} & \cellcolor[HTML]{FFCBCB}{99.17} & \cellcolor[HTML]{E8F0FA}{92.83} & \cellcolor[HTML]{FFE6E6}{90.33} & \cellcolor[HTML]{EBF3FA}{91.92} & \cellcolor[HTML]{FFE7E7}{89.58} & \cellcolor[HTML]{F6FAFD}{78.92} & \cellcolor[HTML]{FFF4F4}{76.08} 
      & \cellcolor[HTML]{F6FAFD}{79.92} & \cellcolor[HTML]{FFF4F4}{77.42} & \cellcolor[HTML]{E0ECF6}{96.08} & \cellcolor[HTML]{FFDADA}{95.67} & \cellcolor[HTML]{BEDBEE}{99.33} & \cellcolor[HTML]{FFCCCC}{98.75} & \cellcolor[HTML]{BEDBEE}{99.58} & \cellcolor[HTML]{FFCBCB}{99.25} & \cellcolor[HTML]{F0F5FC}{87.58} & \cellcolor[HTML]{FFECEC}{84.50} \\

      & \cellcolor[HTML]{D4E6F1}{96.75} & \cellcolor[HTML]{FFD9D9}{96.33} & \cellcolor[HTML]{F0F5FC}{90.00} & \cellcolor[HTML]{FFE9E9}{88.58} & \cellcolor[HTML]{C4DEF0}{99.08} & \cellcolor[HTML]{FFD5D5}{96.17} & \cellcolor[HTML]{CCE2F2}{98.58} & \cellcolor[HTML]{FFDDDD}{93.17} & \cellcolor[HTML]{DCEBF5}{97.33} & \cellcolor[HTML]{FFD9D9}{94.75} 
      & \cellcolor[HTML]{CCE2F2}{98.58} & \cellcolor[HTML]{FFE0E0}{92.25} & \cellcolor[HTML]{E8F0FA}{91.75} & \cellcolor[HTML]{FFE6E6}{91.00} & \cellcolor[HTML]{FAFCFE}{68.17} & \cellcolor[HTML]{FFF6F6}{64.42} & \cellcolor[HTML]{E0ECF6}{95.50} & \cellcolor[HTML]{FFEBEB}{86.67} & \cellcolor[HTML]{CEE3F2}{98.33} & \cellcolor[HTML]{FFE1E1}{92.33} & \multirow{2}{*}{90.30} & \multirow{2}{*}{87.04} \\
      \multirow{-2}{*}{GPT-4o-mini} 
      & \cellcolor[HTML]{CEE3F2}{99.00} & \cellcolor[HTML]{FFD2D2}{98.33} & \cellcolor[HTML]{CEE3F2}{97.50} & \cellcolor[HTML]{FFD5D5}{96.67} & \cellcolor[HTML]{F2F7FD}{85.67} & \cellcolor[HTML]{FFF0F0}{81.58} & \cellcolor[HTML]{F0F5FC}{88.00} & \cellcolor[HTML]{FFECEC}{84.17} & \cellcolor[HTML]{FBFDFE}{67.00} & \cellcolor[HTML]{FFF6F6}{62.25} 
      & \cellcolor[HTML]{F9FBFE}{74.17} & \cellcolor[HTML]{FFF3F3}{71.67} & \cellcolor[HTML]{E4EFF8}{95.08} & \cellcolor[HTML]{FFE2E2}{92.67} & \cellcolor[HTML]{CCE2F2}{98.67} & \cellcolor[HTML]{FFD9D9}{95.75} & \cellcolor[HTML]{EFF5FC}{97.92} & \cellcolor[HTML]{FFD9D9}{96.08} & \cellcolor[HTML]{FAFCFE}{69.00} & \cellcolor[HTML]{FFF5F5}{65.92} \\

      & \cellcolor[HTML]{C0DCEF}{99.25} & \cellcolor[HTML]{FFD5D5}{96.08} & \cellcolor[HTML]{C0DCEF}{99.42} & \cellcolor[HTML]{FFCCCC}{99.25} & \cellcolor[HTML]{BEDBEE}{99.50} & \cellcolor[HTML]{FFCBCB}{99.17} & \cellcolor[HTML]{BEDBEE}{99.75} & \cellcolor[HTML]{FFDDDD}{93.17} & \cellcolor[HTML]{DCEBF5}{97.58} & \cellcolor[HTML]{FFDBDB}{95.75} 
      & \cellcolor[HTML]{C2DDEF}{99.33} & \cellcolor[HTML]{FFE0E0}{92.92} & \cellcolor[HTML]{C2DDEF}{99.33} & \cellcolor[HTML]{FFD5D5}{96.67} & \cellcolor[HTML]{FAFCFE}{88.08} & \cellcolor[HTML]{FFF2F2}{83.58} & \cellcolor[HTML]{CCE2F2}{97.58} & \cellcolor[HTML]{FFEAEA}{93.92} & \cellcolor[HTML]{CEE3F2}{98.83} & \cellcolor[HTML]{FFDEDE}{97.00} & \multirow{2}{*}{96.01} & \multirow{2}{*}{94.36} \\
      \multirow{-2}{*}{\UniTrans{}} 
      & \cellcolor[HTML]{B8D8EC}{100.00} & \cellcolor[HTML]{FFCDCD}{99.83} & \cellcolor[HTML]{B8D8EC}{100.00} & \cellcolor[HTML]{FFCCCC}{99.67} & \cellcolor[HTML]{E0ECF6}{95.50} & \cellcolor[HTML]{FFDFDF}{93.17} & \cellcolor[HTML]{F0F5FC}{93.83} & \cellcolor[HTML]{FFF1F1}{91.17} & \cellcolor[HTML]{E4EFF8}{92.25} & \cellcolor[HTML]{FFE7E7}{89.67} 
      & \cellcolor[HTML]{F6FAFD}{87.58} & \cellcolor[HTML]{FFF1F1}{90.42} & \cellcolor[HTML]{C0DCEF}{99.33} & \cellcolor[HTML]{FFCFCF}{99.00} & \cellcolor[HTML]{C0DCEF}{99.42} & \cellcolor[HTML]{FFD2D2}{98.83} & \cellcolor[HTML]{C0DCEF}{99.75} & \cellcolor[HTML]{FFDADA}{99.33} & \cellcolor[HTML]{F0F5FC}{86.75} & \cellcolor[HTML]{FFF3F3}{82.42} \\

      &       &       & \cellcolor[HTML]{F4F9FE}{81.50} & \cellcolor[HTML]{FFF3F3}{77.58} & \cellcolor[HTML]{C2DDEF}{98.83} & \cellcolor[HTML]{FFDDDD}{94.08} &       &       & \cellcolor[HTML]{D6E7F3}{96.33} & \cellcolor[HTML]{FFE1E1}{92.25}
      &       &       &       &       &       &       &       &       &       &       & \multirow{2}{*}{86.71} & \multirow{2}{*}{81.58} \\
      \multirow{-2}{*}{ExeCoder}
      &       &       & \cellcolor[HTML]{F2F7FD}{92.92} & \cellcolor[HTML]{FFE7E7}{90.83} & \cellcolor[HTML]{F0F5FC}{82.92} & \cellcolor[HTML]{FFF5F5}{74.92} &       &       & \cellcolor[HTML]{FAFCFE}{67.75} & \cellcolor[HTML]{FFF7F7}{59.83}
      &       &       &       &       &       &       &       &       &       &       \\

      & \cellcolor[HTML]{D0E4F2}{97.08} & \cellcolor[HTML]{FFDADA}{95.42} & \cellcolor[HTML]{E4EFF8}{94.33} & \cellcolor[HTML]{FFE6E6}{91.17} & \cellcolor[HTML]{E8F0FA}{92.75} & \cellcolor[HTML]{FFEBEB}{87.25} & \cellcolor[HTML]{CEE3F2}{97.50} & \cellcolor[HTML]{FFE3E3}{90.50} & \cellcolor[HTML]{DCEBF5}{97.67} & \cellcolor[HTML]{FFE0E0}{92.25} 
      & \cellcolor[HTML]{D0E4F2}{97.25} & \cellcolor[HTML]{FFE3E3}{90.42} & \cellcolor[HTML]{E8F0FA}{94.00} & \cellcolor[HTML]{FFE5E5}{91.58} & \cellcolor[HTML]{F2F7FD}{83.08} & \cellcolor[HTML]{FFF1F1}{74.83} & \cellcolor[HTML]{E4EFF8}{94.25} & \cellcolor[HTML]{FFF0F0}{80.75} & \cellcolor[HTML]{D4E6F1}{97.33} & \cellcolor[HTML]{FFE3E3}{90.58} & \multirow{2}{*}{91.48} & \multirow{2}{*}{85.90} \\
      \multirow{-2}{*}{IT(Magi)} 
      & \cellcolor[HTML]{E4EFF8}{94.50} & \cellcolor[HTML]{FFE2E2}{92.17} & \cellcolor[HTML]{E6F0F9}{93.00} & \cellcolor[HTML]{FFE6E6}{90.83} & \cellcolor[HTML]{E8F0FA}{92.75} & \cellcolor[HTML]{FFECEC}{85.58} & \cellcolor[HTML]{E8F0FA}{93.75} & \cellcolor[HTML]{FFE7E7}{89.17} & \cellcolor[HTML]{F8FBFE}{77.00} & \cellcolor[HTML]{FFF4F4}{68.67} 
      & \cellcolor[HTML]{F9FBFE}{75.75} & \cellcolor[HTML]{FFF4F4}{68.67} & \cellcolor[HTML]{F0F5FC}{84.75} & \cellcolor[HTML]{FFF0F0}{81.50} & \cellcolor[HTML]{CCE2F2}{98.50} & \cellcolor[HTML]{FFD9D9}{93.75} & \cellcolor[HTML]{E6F0F9}{93.33} & \cellcolor[HTML]{FFEBEB}{89.00} & \cellcolor[HTML]{F6FAFD}{81.00} & \cellcolor[HTML]{FFF2F2}{73.92} \\

      & \cellcolor[HTML]{FAFCFE}{66.25} & \cellcolor[HTML]{FFF5F5}{65.17} & \cellcolor[HTML]{F9FBFE}{70.08} & \cellcolor[HTML]{FFF4F4}{66.67} & \cellcolor[HTML]{F6FAFD}{75.67} & \cellcolor[HTML]{FFF3F3}{66.83} & \cellcolor[HTML]{F8FBFE}{69.50} & \cellcolor[HTML]{FFF6F6}{58.17} & \cellcolor[HTML]{FCFDFE}{61.50} & \cellcolor[HTML]{FFF7F7}{54.92} 
      & \cellcolor[HTML]{FCFDFE}{59.58} & \cellcolor[HTML]{FFFFF8}{48.08} & \cellcolor[HTML]{FCFDFE}{59.75} & \cellcolor[HTML]{FFF6F6}{58.08} & \cellcolor[HTML]{FEFEFE}{55.08} & \cellcolor[HTML]{FFF7F7}{51.08} & \cellcolor[HTML]{FAFCFE}{66.83} & \cellcolor[HTML]{FFF6F6}{53.42} & \cellcolor[HTML]{FCFDFE}{61.58} & \cellcolor[HTML]{FFFFF9}{48.83} & \multirow{2}{*}{65.17} & \multirow{2}{*}{59.30} \\
      \multirow{-2}{*}{IT(Star)}
      & \cellcolor[HTML]{F6FAFD}{73.08} & \cellcolor[HTML]{FFF3F3}{71.67} & \cellcolor[HTML]{F4F9FE}{81.17} & \cellcolor[HTML]{FFEEEE}{78.83} & \cellcolor[HTML]{F6FAFD}{78.67} & \cellcolor[HTML]{FFF2F2}{70.92} & \cellcolor[HTML]{F4F9FE}{81.17} & \cellcolor[HTML]{FFEFEF}{77.08} & \cellcolor[HTML]{FEFEFE}{51.67} & \cellcolor[HTML]{FFFFF9}{45.17} 
      & \cellcolor[HTML]{FEFEFE}{51.75} & \cellcolor[HTML]{FFFFF9}{47.00} & \cellcolor[HTML]{FEFEFE}{48.33} & \cellcolor[HTML]{FFFFF9}{46.00} & \cellcolor[HTML]{FAFCFE}{68.58} & \cellcolor[HTML]{FFF6F6}{60.67} & \cellcolor[HTML]{F9FBFE}{70.83} & \cellcolor[HTML]{FFF4F4}{67.42} & \cellcolor[HTML]{FEFEFE}{52.42} & \cellcolor[HTML]{FFFFF9}{50.00} \\

      & \cellcolor[HTML]{FEFEFE}{24.25} & \cellcolor[HTML]{FFFFFB}{20.50} &       &       &       &       &       &       &       &       
      &       &       &       &       &       &       &       &       &       &       & \multirow{2}{*}{22.88} & \multirow{2}{*}{19.25} \\
      \multirow{-2}{*}{StructCoder}
      & \cellcolor[HTML]{FEFEFE}{21.50} & \cellcolor[HTML]{FFFFFC}{18.00} &       &       &       &       &       &       &       &       
      &       &       &       &       &       &       &       &       &       &  \\

      & \cellcolor[HTML]{FBFDFE}{56.25} & \cellcolor[HTML]{FFF7F7}{52.00} & \cellcolor[HTML]{FCFDFE}{52.25} & \cellcolor[HTML]{FFF8F8}{48.00} & \cellcolor[HTML]{FCFDFE}{51.25} & \cellcolor[HTML]{FFFFF4}{21.75} & \cellcolor[HTML]{FAFCFE}{66.50} & \cellcolor[HTML]{FFF6F6}{43.75} & \cellcolor[HTML]{FCFDFE}{47.25} & \cellcolor[HTML]{FFFFF6}{17.75} 
      & \cellcolor[HTML]{FCFDFE}{61.75} & \cellcolor[HTML]{FFF9F9}{42.25} & \cellcolor[HTML]{FCFDFE}{49.00} & \cellcolor[HTML]{FFF7F7}{44.25} & \cellcolor[HTML]{FEFEFE}{33.50} & \cellcolor[HTML]{FFFFFB}{24.00} & \cellcolor[HTML]{FEFEFE}{50.50} & \cellcolor[HTML]{FFFFF8}{26.75} & \cellcolor[HTML]{FCFDFE}{63.50} & \cellcolor[HTML]{FFF7F7}{43.25} & \multirow{2}{*}{48.10} & \multirow{2}{*}{34.30} \\
      \multirow{-2}{*}{VIM-PT}
      & \cellcolor[HTML]{FEFEFE}{42.25} & \cellcolor[HTML]{FFFFFA}{40.75} & \cellcolor[HTML]{FEFEFE}{44.75} & \cellcolor[HTML]{FFFFFB}{39.75} & \cellcolor[HTML]{FEFEFE}{31.00} & \cellcolor[HTML]{FFFFF9}{24.75} & \cellcolor[HTML]{FEFEFE}{40.00} & \cellcolor[HTML]{FFFFFA}{35.50} & \cellcolor[HTML]{FEFEFE}{29.25} & \cellcolor[HTML]{FFFFF9}{19.25} 
      & \cellcolor[HTML]{FEFEFE}{46.00} & \cellcolor[HTML]{FFFFFA}{40.75} & \cellcolor[HTML]{FCFDFE}{53.25} & \cellcolor[HTML]{FFF8F8}{48.00} & \cellcolor[HTML]{FEFEFE}{50.75} & \cellcolor[HTML]{FFFFF6}{20.00} & \cellcolor[HTML]{FEFEFE}{50.00} & \cellcolor[HTML]{FFFFF8}{13.50} & \cellcolor[HTML]{FEFEFE}{43.00} & \cellcolor[HTML]{FFFFFB}{40.00} \\

      &       &       & \cellcolor[HTML]{F6FAFD}{76.00} & \cellcolor[HTML]{FFF2F2}{64.75} & \cellcolor[HTML]{F2F7FD}{84.00} & \cellcolor[HTML]{FFEFEF}{70.25} &       &       & \cellcolor[HTML]{F9FBFE}{71.00} & \cellcolor[HTML]{FFF4F4}{60.50} 
      &       &       &       &       &       &       &       &       &       &       & \multirow{2}{*}{70.50} & \multirow{2}{*}{56.96} \\
      \multirow{-2}{*}{TransCoder-ST}
      &       &       & \cellcolor[HTML]{F8FBFE}{68.75} & \cellcolor[HTML]{FFF6F6}{58.00} & \cellcolor[HTML]{F9FBFE}{70.00} & \cellcolor[HTML]{FFF7F7}{52.50} &       &       & \cellcolor[HTML]{FEFEFE}{53.25} & \cellcolor[HTML]{FFFFFA}{35.75} 
      &       &       &       &       &       &       &       &       &       &  \\

      & \cellcolor[HTML]{F6FAFD}{75.00} & \cellcolor[HTML]{FFF4F4}{67.25} & \cellcolor[HTML]{FCFDFE}{62.75} & \cellcolor[HTML]{FFF7F7}{54.00} & \cellcolor[HTML]{F2F7FD}{84.25} & \cellcolor[HTML]{FFF1F1}{63.75} & \cellcolor[HTML]{F4F9FE}{81.75} & \cellcolor[HTML]{FFF1F1}{63.00} & \cellcolor[HTML]{F2F7FD}{83.25} & \cellcolor[HTML]{FFEFEF}{66.50} 
      & \cellcolor[HTML]{F6FAFD}{79.50} & \cellcolor[HTML]{FFF3F3}{59.75} & \cellcolor[HTML]{F9FBFE}{71.50} & \cellcolor[HTML]{FFF2F2}{64.25} & \cellcolor[HTML]{FCFDFE}{60.25} & \cellcolor[HTML]{FFF7F7}{51.25} & \cellcolor[HTML]{F6FAFD}{80.25} & \cellcolor[HTML]{FFF7F7}{56.50} & \cellcolor[HTML]{F6FAFD}{79.00} & \cellcolor[HTML]{FFF4F4}{59.25} & \multirow{2}{*}{68.80} & \multirow{2}{*}{56.06} \\
      \multirow{-2}{*}{CodeT5+}
      & \cellcolor[HTML]{FEFEFE}{57.75} & \cellcolor[HTML]{FFFFF9}{53.25} & \cellcolor[HTML]{FCFDFE}{60.50} & \cellcolor[HTML]{FFF8F8}{55.25} & \cellcolor[HTML]{FEFEFE}{55.00} & \cellcolor[HTML]{FFFFF9}{45.50} & \cellcolor[HTML]{FEFEFE}{55.50} & \cellcolor[HTML]{FFFFF9}{48.50} & \cellcolor[HTML]{FEFEFE}{50.75} & \cellcolor[HTML]{FFFFFA}{39.50} 
      & \cellcolor[HTML]{FEFEFE}{52.25} & \cellcolor[HTML]{FFFFFA}{44.00} & \cellcolor[HTML]{FEFEFE}{58.00} & \cellcolor[HTML]{FFFFF9}{49.00} & \cellcolor[HTML]{F2F7FD}{86.25} & \cellcolor[HTML]{FFEFEF}{65.00} & \cellcolor[HTML]{F4F9FE}{81.00} & \cellcolor[HTML]{FFF1F1}{63.00} & \cellcolor[HTML]{FCFDFE}{61.50} & \cellcolor[HTML]{FFF7F7}{52.75} \\
    
    \bottomrule
    \end{tabular}
    \end{threeparttable}
\end{table*}
\begin{table}[tbp]
  \centering
  \footnotesize
  \caption{Translation performance on TransCoder-Uni.}
  \label{tab:overall-transuni}
  \resizebox{\columnwidth}{!}{
  \begin{threeparttable}
  \setlength{\tabcolsep}{4pt} 
  \begin{tabular}{l *{6}{c} cc}
    \toprule
    \multirow{2}{*}{Name} 
    & \multicolumn{2}{c}{Java$\rightleftharpoons$C++} 
    & \multicolumn{2}{c}{Java$\rightleftharpoons$Py.} 
    & \multicolumn{2}{c}{C++$\rightleftharpoons$Py.} 
    & \multicolumn{2}{c}{Average} \\
    \cmidrule(lr){2-3} \cmidrule(lr){4-5} \cmidrule(lr){6-7} \cmidrule(lr){8-9}
    & CSR & CA & CSR & CA & CSR & CA & CSR & CA \\
    \midrule

      & \cellcolor[HTML]{D4E6F1}{97.57} & \cellcolor[HTML]{FFD7D7}{94.93} & \cellcolor[HTML]{D4E6F1}{97.49} & \cellcolor[HTML]{FFE0E0}{91.31} & \cellcolor[HTML]{DBEAF4}{96.34} & \cellcolor[HTML]{FFE7E7}{89.30} & \multirow{2}{*}{94.56} & \multirow{2}{*}{89.79} \\
    \multirow{-2}{*}{DS-V4-Flash} 
      & \cellcolor[HTML]{E8F0FA}{94.95} & \cellcolor[HTML]{FFDEDE}{92.05} & \cellcolor[HTML]{F0F5FC}{92.60} & \cellcolor[HTML]{FFECEC}{88.38} & \cellcolor[HTML]{F8FBFE}{88.51} & \cellcolor[HTML]{FFF2F2}{82.73} & & \\

      & \cellcolor[HTML]{D4E6F1}{97.29} & \cellcolor[HTML]{FFD6D6}{94.00} & \cellcolor[HTML]{D4E6F1}{97.13} & \cellcolor[HTML]{FFE3E3}{90.80} & \cellcolor[HTML]{E0ECF6}{95.11} & \cellcolor[HTML]{FFEEEE}{86.42} & \multirow{2}{*}{95.15} & \multirow{2}{*}{90.09} \\
    \multirow{-2}{*}{GPT-3.5-turbo} 
      & \cellcolor[HTML]{E8F0FA}{94.33} & \cellcolor[HTML]{FFE4E4}{91.70} & \cellcolor[HTML]{ECF4FB}{91.77} & \cellcolor[HTML]{FFE8E8}{87.34} & \cellcolor[HTML]{E0ECF6}{95.43} & \cellcolor[HTML]{FFE8E8}{90.29} & & \\

      & \cellcolor[HTML]{D4E6F1}{97.29} & \cellcolor[HTML]{FFDEDE}{92.43} & \cellcolor[HTML]{E0ECF6}{94.68} & \cellcolor[HTML]{FFE6E6}{87.72} & \cellcolor[HTML]{E6F0F9}{93.89} & \cellcolor[HTML]{FFEEEE}{85.42} & \multirow{2}{*}{91.84} & \multirow{2}{*}{86.31} \\
    \multirow{-2}{*}{GPT-4o-mini} 
      & \cellcolor[HTML]{F2F7FD}{89.14} & \cellcolor[HTML]{FFF0F0}{86.72} & \cellcolor[HTML]{F9FBFE}{84.99} & \cellcolor[HTML]{FFF6F6}{80.15} & \cellcolor[HTML]{EFF5FC}{91.36} & \cellcolor[HTML]{FFEEEE}{85.58} & & \\

      & \cellcolor[HTML]{C9E0EF}{97.86} & \cellcolor[HTML]{FFD2D2}{94.86} & \cellcolor[HTML]{C9E0EF}{97.84} & \cellcolor[HTML]{FFE1E1}{91.52} & \cellcolor[HTML]{DBEAF4}{96.55} & \cellcolor[HTML]{FFE6E6}{88.29} & \multirow{2}{*}{96.59} & \multirow{2}{*}{91.66} \\
    \multirow{-2}{*}{UniTrans} 
      & \cellcolor[HTML]{E4EFF8}{95.30} & \cellcolor[HTML]{FFE0E0}{92.74} & \cellcolor[HTML]{E4EFF8}{94.61} & \cellcolor[HTML]{FFE6E6}{90.46} & \cellcolor[HTML]{CEE3F2}{97.50} & \cellcolor[HTML]{FFDFDF}{92.08} & & \\

      & \cellcolor[HTML]{F0F5FC}{91.94} & \cellcolor[HTML]{FFECEC}{87.22} & \cellcolor[HTML]{E6F0F9}{94.68} & \cellcolor[HTML]{FFECEC}{87.14} & \cellcolor[HTML]{E6F0F9}{94.75} & \cellcolor[HTML]{FFECEC}{86.78} & \multirow{2}{*}{94.49} & \multirow{2}{*}{87.41} \\
    \multirow{-2}{*}{ExeCoder} 
      & \cellcolor[HTML]{BEDBEE}{99.59} & \cellcolor[HTML]{FFD2D2}{96.61} & \cellcolor[HTML]{C9E0EF}{97.93} & \cellcolor[HTML]{FFE9E9}{88.73} & \cellcolor[HTML]{F8FBFE}{88.08} & \cellcolor[HTML]{FFF4F4}{77.94} & & \\

      & \cellcolor[HTML]{D4E6F1}{96.93} & \cellcolor[HTML]{FFE0E0}{92.29} & \cellcolor[HTML]{EFF5FC}{91.09} & \cellcolor[HTML]{FFF0F0}{83.05} & \cellcolor[HTML]{E4EFF8}{92.24} & \cellcolor[HTML]{FFF2F2}{81.68} & \multirow{2}{*}{94.22} & \multirow{2}{*}{87.23} \\
    \multirow{-2}{*}{IT(Magi)} 
      & \cellcolor[HTML]{E0ECF6}{95.71} & \cellcolor[HTML]{FFE1E1}{92.12} & \cellcolor[HTML]{E8F0FA}{93.98} & \cellcolor[HTML]{FFECEC}{87.00} & \cellcolor[HTML]{E0ECF6}{95.29} & \cellcolor[HTML]{FFECEC}{87.01} & & \\

      & \cellcolor[HTML]{FCFDFE}{62.38} & \cellcolor[HTML]{FFFDFD}{59.39} & \cellcolor[HTML]{FEFEFE}{57.26} & \cellcolor[HTML]{FFFEFE}{50.14} & \cellcolor[HTML]{FEFEFE}{53.66} & \cellcolor[HTML]{FFFEFE}{46.34} & \multirow{2}{*}{61.58} & \multirow{2}{*}{55.73} \\
    \multirow{-2}{*}{IT(Star)} 
      & \cellcolor[HTML]{FEFEFE}{52.56} & \cellcolor[HTML]{FFFEFE}{50.69} & \cellcolor[HTML]{FCFDFE}{66.74} & \cellcolor[HTML]{FFF9F9}{61.27} & \cellcolor[HTML]{F6FAFD}{76.95} & \cellcolor[HTML]{FFF6F6}{66.45} & & \\

      & \cellcolor[HTML]{FBFDFE}{76.02} & \cellcolor[HTML]{FFF5F5}{68.95} & \cellcolor[HTML]{FEFEFE}{59.05} & \cellcolor[HTML]{FFFFF7}{29.31} & \cellcolor[HTML]{FEFEFE}{55.17} & \cellcolor[HTML]{FFFFF8}{25.43} & \multirow{2}{*}{61.46} & \multirow{2}{*}{45.33} \\
    \multirow{-2}{*}{VIM-PT} 
      & \cellcolor[HTML]{FBFDFE}{71.16} & \cellcolor[HTML]{FFF7F7}{66.18} & \cellcolor[HTML]{FEFEFE}{52.07} & \cellcolor[HTML]{FFFFFB}{42.95} & \cellcolor[HTML]{FEFEFE}{55.25} & \cellcolor[HTML]{FFFFFB}{38.33} & & \\

      & \cellcolor[HTML]{F1F6FD}{92.72} & \cellcolor[HTML]{FFECEC}{85.01} & \cellcolor[HTML]{F7FAFE}{78.88} & \cellcolor[HTML]{FFF4F4}{69.18} & \cellcolor[HTML]{FAFCFE}{72.41} & \cellcolor[HTML]{FFF6F6}{62.93} & \multirow{2}{*}{78.03} & \multirow{2}{*}{68.75} \\
    \multirow{-2}{*}{Transcoder-ST} 
      & \cellcolor[HTML]{F8FBFE}{75.10} & \cellcolor[HTML]{FFF2F2}{70.95} & \cellcolor[HTML]{F8FBFE}{72.61} & \cellcolor[HTML]{FFF6F6}{62.24} & \cellcolor[HTML]{F6FAFD}{76.66} & \cellcolor[HTML]{FFF6F6}{62.10} & & \\

      & \cellcolor[HTML]{F0F5FC}{85.65} & \cellcolor[HTML]{FFF0F0}{75.16} & \cellcolor[HTML]{F2F7FD}{83.62} & \cellcolor[HTML]{FFF2F2}{71.77} & \cellcolor[HTML]{F2F7FD}{82.97} & \cellcolor[HTML]{FFF2F2}{72.20} & \multirow{2}{*}{81.85} & \multirow{2}{*}{73.07} \\
    \multirow{-2}{*}{CodeT5+} 
      & \cellcolor[HTML]{F4F9FE}{81.33} & \cellcolor[HTML]{FFF3F3}{78.63} & \cellcolor[HTML]{F9FBFE}{75.31} & \cellcolor[HTML]{FFF4F4}{70.95} & \cellcolor[HTML]{F4F9FE}{82.44} & \cellcolor[HTML]{FFF5F5}{69.59} & & \\

    \bottomrule
  \end{tabular}
  \end{threeparttable}
  }
\end{table}

In terms of correctness metrics, LLM-based methods, including general-purpose LLMs and methods enhanced for code translation, generally outperform state-of-the-art learning-based methods across metrics and datasets. Specifically, except for \ITStar{}, most LLM-based methods maintain high average CA and CSR scores on both datasets, clearly outperforming traditional learning-based methods. In contrast, the best-performing learning-based method, \TranscoderST{}, achieves CA and CSR scores of only 56.96\% and 70.50\%, respectively, which are much lower than those of most LLM-based methods.

Overall, among all translation tasks, UniTrans achieves the best performance, with average CSR (CA) scores of 96.01\% (94.36\%) on G-TransEval and 96.59\% (91.66\%) on TransCoder-Uni, significantly outperforming not only learning-based methods but also other LLM-based approaches. Among all methods that enhance performance using large language models, UniTrans improves the CSR (CA) over its base model (GPT-3.5-turbo) by 1.18 percentage points (1.13 percentage points) and 1.44 percentage points (1.57 percentage points) on the two datasets, respectively, while ExeCoder and IT (Magi) achieve performance close to that of GPT-4o-mini.
This suggests, to some extent, that current LLM methods designed for enhancing code translation tasks generally achieve stable and relatively high correctness, but the extent of improvement over the base LLM still depends on the specific method design, the capability of the base model, and the particular task at hand.

\begin{finding}
In method-level code translation tasks, LLMs and LLM-based methods significantly outperform learning-based methods in correctness metrics. Methods enhanced for code translation usually achieve strong performance, but their improvements over base LLMs depend on the specific method design, the capability of the base model, and the target tasks.
\end{finding}

In terms of similarity metrics, almost all methods achieve average CodeBLEU scores above 70.00\% on both datasets. However, the advantage of LLM-based methods on this metric is less pronounced than on correctness metrics. Specifically, although learning-based methods usually obtain lower CA scores, their similarity scores are close to those of LLM-based methods. For example, \StructCoder{} and \VIMPT{} perform poorly in correctness metrics, but their similarity scores are close to those of \UniTrans{}. This indicates that learning-based methods often generate outputs that are superficially similar to the reference translations but fail to compile or pass tests.

\begin{figure}[tbp]
\centering
\includegraphics[width=0.97\columnwidth]{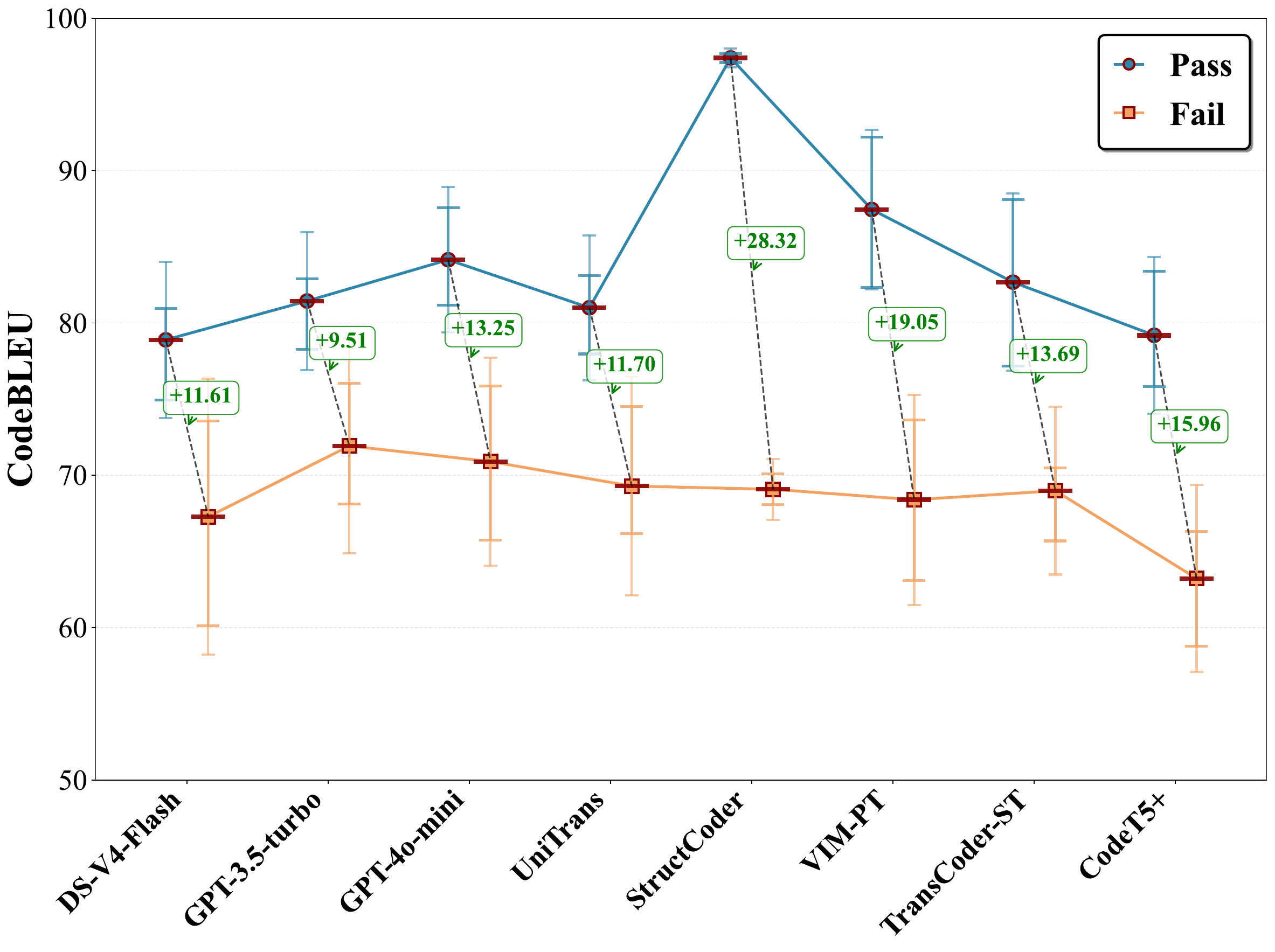}
\caption{CodeBLEU gaps between passed and failed samples. }
\label{figs:pass-fail}
\end{figure}

To further investigate this phenomenon, we group all passed and failed outputs generated by different methods and compute their CodeBLEU scores. We exclude \IT{} to avoid the impact of empty parser outputs on the calculation. As shown in Figure~\ref{figs:pass-fail}, the average CodeBLEU gap between passed and failed samples is 15.96\% to 28.32\% for learning-based methods, while it is only 9.51\% to 13.25\% for LLM-based methods, which is substantially smaller. This suggests that LLM-based methods tend to generate functionally correct outputs even when their surface similarity to the reference is relatively low, indicating that they focus more on understanding the semantic functionality of source programs rather than performing line-by-line translation. Combining correctness and similarity analyses, we conclude that recent advances in LLMs have dominant effectiveness in code translation: they consistently outperform learning-based methods, provide more robust translation results, and emphasize functional equivalence rather than surface similarity.

\begin{finding}
Learning-based methods often generate outputs that are similar to correct translations but cannot compile or pass tests. In contrast, LLM-based methods achieve CodeBLEU scores comparable to learning-based methods and show a smaller CodeBLEU gap between correct and incorrect outputs, indicating that they focus more on semantic functionality rather than line-by-line translation.
\end{finding}

\subsection{RQ2: Performance on Different Translation Tasks}

According to the results shown in Table~\ref{tab:overall-gtrans} and Table~\ref{tab:overall-transuni}, the same method may achieve substantially different results when translating between the same pair of programming languages in opposite directions. For example, when using DeepSeek-V4-Flash on the G-TransEval dataset, the CA score for C++ to Python reaches 99.25\%, whereas the score in the reverse direction drops sharply to 76.33\%. Similar patterns can also be observed for other large language models and LLM-based methods. To better understand this issue, we conduct a deeper analysis of the results for each programming language. Specifically, we compute the average CA when each language is used as the source or target language on the two datasets. The results are shown in Table~\ref{tab:merged-source} and Table~\ref{tab:merged-target}.

\begin{table}[tbp]
  \centering
  \caption{Average CA for translation from each source language (presented in the table) to all other target languages.}
  \label{tab:merged-source}
  \resizebox{\columnwidth}{!}{
  \begin{tabular}{lcccccccc}
    \toprule
    \multirow{2}{*}{Model} & 
    \multicolumn{5}{c}{G-TransEval} & 
    \multicolumn{3}{c}{TransCoder-Uni} \\
    \cmidrule(lr){2-6} \cmidrule(l){7-9}
     & Java & C++ & Python & C\# & JS & Java & C++ & Python \\
    \midrule
    DS-V4-Flash   & 97.15 & 98.50 & 87.08 & 97.08 & 84.50 & 93.13 & 90.70 & 85.60 \\
    GPT-3.5-turbo   & 97.31 & 97.55 & 86.00 & 97.63 & 87.69 & 92.41 & 89.11 & 88.80 \\
    GPT-4o-mini     & 93.56 & 93.67 & 73.73 & 94.77 & 79.46 & 90.08 & 86.08 & 82.82 \\
    \UniTrans{}     & 98.00 & 97.77 & 87.23 & 98.66 & 90.12 & 91.60 & 93.47 & 89.91 \\
    \ExeCoder{}     &85.83 & 91.54 & 67.38 & /     & /     & 87.18 & 91.70 & 83.34 \\
    \ITMagi{}       & 91.09 & 91.27 & 77.46 & 89.50 & 80.19 & 87.68 & 87.00 & 87.00 \\
    \ITStar{}       & 64.21 & 59.98 & 55.15 & 56.79 & 60.38 & 54.78 & 48.56 & 63.82 \\
    \StructCoder{}     & 20.50 & /     & /     & 18.00 & /     & /    & /    & /    \\
    \VIMPT{}          & 41.38 & 36.00 & 23.69 & 38.00 & 32.44 & 49.19 & 46.19 & 40.67 \\
    \TranscoderST{}   & 67.50 & 59.25 & 44.12 & /     & /     & 77.12 & 67.02 & 62.28 \\
    CodeT5+         & 62.00 & 61.44 & 48.19 & 56.62 & 52.06 & 73.47 & 75.48 & 70.28 \\
    \bottomrule
  \end{tabular}
  }
\end{table}
\begin{table}[tbp]
  \centering
  \caption{Average CA for translation from all
other languages to each target language (presented in the table).}
  \label{tab:merged-target}
  \resizebox{\columnwidth}{!}{
  \begin{tabular}{lcccccccc}
    \toprule
    \multirow{2}{*}{Model} & 
    \multicolumn{5}{c}{G-TransEval} & 
    \multicolumn{3}{c}{TransCoder-Uni} \\
    \cmidrule(lr){2-6} \cmidrule(l){7-9}
     & Java & C++ & Python & C\# & JS & Java & C++ & Python \\
    \midrule
    DS-V4-Flash   & 95.09 & 84.00 & 99.35 & 89.84 & 96.04 & 90.21 & 88.83 & 90.30 \\
    GPT-3.5-turbo   & 94.69 & 85.86 & 98.27 & 91.92 & 95.44 & 89.52 & 92.15 & 88.61 \\
    GPT-4o-mini     & 90.19 & 78.79 & 95.69 & 79.42 & 91.11 & 83.44 & 89.01 & 86.57 \\
    \UniTrans{}     & 95.38 & 90.66 & 98.31 & 90.94 & 96.50 & 93.20 & 90.56 & 91.25 \\
    \ExeCoder{}     &82.88 & 68.71 & 93.17 & /     & /     & 92.67 & 82.58 & 86.96 \\
    \ITMagi{}  & 89.44 & 77.50 & 90.56 & 83.94 & 88.06 & 89.56 & 89.65 & 82.36 \\
    \ITStar{}  & 74.62 & 51.21 & 62.46 & 56.08 & 52.12 & 55.98 & 62.92 & 48.24 \\
    \StructCoder{}     & 18.00 & /     & /     & 20.50 & /     & /     & /     & /     \\
    \VIMPT{}          & 35.19 & 39.00 & 18.25 & 40.06 & 39.00 & 54.56 & 53.64 & 27.37 \\
    \TranscoderST{}   & 55.25 & 50.25 & 65.38 & /     & /     & 66.60 & 73.66 & 66.06 \\
    CodeT5+         & 50.62 & 46.62 & 64.56 & 58.88 & 59.62 & 74.79 & 72.38 & 71.98 \\
    \bottomrule
  \end{tabular}
  }
\end{table}

The results confirm that the choice of source and target languages has a clear impact. On the G-TransEval dataset, we observe a phenomenon consistent with prior studies~\cite{polyhumaneval, classeval}: translation performs better when Python is used as the target language. We further extend and summarize this observation. Specifically, when Python or JavaScript is used as the target language, CA scores are significantly higher than when Java or C++ is used as the target language. Conversely, when Java or C++ is used as the source language, CA scores are significantly higher than when Python or JavaScript is used as the source language. We attribute this to the difference in the amount of explicit information encoded in different programming languages. For example, when translating from Python, where type information is relatively implicit, to Java or C++, where type information is more explicit, the model needs to infer variable types, parameter types, and return types, which may introduce errors.

However, on the TransCoder-Uni dataset, the performance decrease or improvement caused by using Python as the source or target language is much less pronounced. This phenomenon motivates us to investigate the differences between the two datasets. Based on our study of failure cases, we speculate that this is because TransCoder-Uni contains fewer type annotations and more complex logic than G-TransEval, as shown in Table~\ref{tab:ccn} and further discussed in RQ4. This characteristic weakens the disadvantage or advantage usually associated with using Python as the source or target language. Overall, programming languages do affect translation correctness, but the extent of this effect also depends on the type complexity and logic complexity of the dataset.

\begin{finding}
\label{finding:rq2-language}
Translation generally tends to perform better when translating from statically typed languages to dynamically typed languages, whereas the opposite direction results in notably lower performance. This suggests that, in practical code translation, using a statically (dynamically) typed language as the source (target) language is usually a more favorable setting.
\end{finding}

\begin{figure*}[htbp]
\centering
\includegraphics[width=0.98\textwidth]{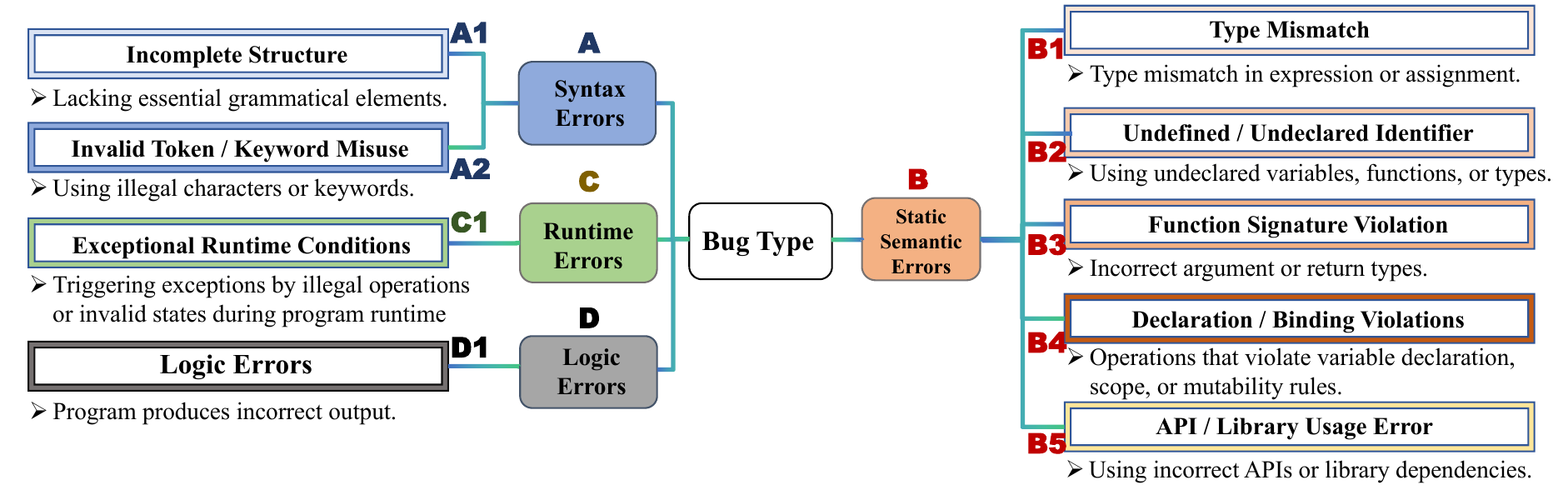}
\caption{Categories of failures caused by the code translation methods studied in our experiments.}
\label{figs/type1}
\end{figure*}

\subsection{RQ3: Class-Level Code Translation Performance}

To further evaluate the practicality of existing code translation techniques in real-world software engineering scenarios, we extend our study from method-level translation to class-level translation. Compared with method-level translation, class-level translation involves longer contexts, richer inter-method dependencies, and more complex type structures. These characteristics make class-level translation not merely a scaled-up version of method-level translation, but a qualitatively more difficult task that requires models and tools to preserve global class semantics, member relationships, and cross-language type consistency.

Table~\ref{tab:classeval} summarizes the class-level translation results in terms of CA and CSR. Overall, all methods show a clear performance decline on class-level translation, especially in CA. Although some methods still achieve relatively high CSR, their CA remains low, indicating that generating compilable class-level code is much easier than generating semantically correct class-level translations.

From the overall averages, the best-performing \UniTrans{} achieves an average CA of 24.82\% and an average CSR of 89.39\%, while the three general-purpose LLMs obtain average CA values between 18.82\% and 22.87\% and average CSR values between 72.15\% and 83.51\%. This result indicates that even the best-performing method can correctly translate only a limited portion of class-level code. In other words, existing techniques still cannot reliably solve class-level code translation in practical settings.

\begin{finding}
Compared with method-level code translation, both LLMs and LLM-based methods show decreases in CA and CSR on class-level code translation. However, the decrease in CSR is much smaller than that in CA, indicating that the main challenge in current class-level code translation is not compilation, but how to generate semantically correct code under more complex code contexts.
\end{finding}

A more detailed analysis shows that, compared with method-level translation, the gap between CA and CSR becomes much larger for almost all methods and translation directions. Specifically, in \texttt{Java$\rightarrow$C++} and \texttt{Python$\rightarrow$C++}, all methods produce extremely low CA values, with several settings close to or equal to zero, while CSR remains moderate or even high. Similarly, for \texttt{C++$\rightarrow$Java} and \texttt{Python$\rightarrow$Java}, CSR is generally above 73.40\%, but CA remains limited ($\leq 25.53\%$). This discrepancy indicates that many translated classes can compile, but still fail to preserve the original program semantics, class structure, or behavioral equivalence. This further suggests that existing methods can handle syntactic issues in translation relatively well, but when source code becomes longer and more complex, involving more dependencies and types, they struggle to achieve the high correctness observed in method-level translation.

The results also reveal a clear language effect in class-level translation. Translations targeting Python achieve relatively higher CA than those targeting Java or C++. In contrast, translating into C++ is consistently the most challenging direction. For example, all methods obtain significantly low CA ($\leq 12.77\%$) on \texttt{Java$\rightarrow$C++}, and only marginally better results on \texttt{Python$\rightarrow$C++}. This trend indicates that when class structures become larger and more interdependent, translating language-specific type systems, object models, and declaration conventions becomes more difficult. Compared with method-level tasks, class-level tasks are more sensitive to language differences, especially when the target language is statically typed and imposes stricter structural and type constraints.

\begin{finding}
Language effects become stronger in class-level code translation. As the translation unit becomes larger and type interactions become more complex, accurately understanding and translating cross-language type systems becomes a major bottleneck. Therefore, practical class-level code translation should place greater emphasis on type-aware translation, especially for tasks targeting statically typed languages, where strict type and structural constraints make translation more difficult.
\end{finding}

\begin{table}[tbp]
  \centering
  \caption{CSR and CA results across different translation directions on ClassEval-T.}
  \label{tab:classeval}
  \resizebox{\columnwidth}{!}{
  \begin{threeparttable}
  \begin{tabular}{l *{5}{cc}}
    \toprule
    \multirow{2}{*}{Task}
    & \multicolumn{2}{c}{ExeCoder}
    & \multicolumn{2}{c}{UniTrans}
    & \multicolumn{2}{c}{DS-V4-Flash}
    & \multicolumn{2}{c}{GPT-3.5-turbo}
    & \multicolumn{2}{c}{GPT-4o-mini} \\
    \cmidrule(lr){2-3}\cmidrule(lr){4-5}\cmidrule(lr){6-7}\cmidrule(lr){8-9}\cmidrule(lr){10-11}
    & \textbf{CSR} & \textbf{CA}
    & \textbf{CSR} & \textbf{CA}
    & \textbf{CSR} & \textbf{CA}
    & \textbf{CSR} & \textbf{CA}
    & \textbf{CSR} & \textbf{CA} \\
    \midrule
    C++ $\rightarrow$ Py.
      & \cellcolor[HTML]{F8FBFE}46.81 & \cellcolor[HTML]{FFF2F2}23.40
      & \cellcolor[HTML]{E8F0FA}79.79 & \cellcolor[HTML]{FFD2D2}54.26
      & \cellcolor[HTML]{F2F7FD}68.09 & \cellcolor[HTML]{FFE0E0}42.55
      & \cellcolor[HTML]{F2F7FD}69.15 & \cellcolor[HTML]{FFE0E0}42.55
      & \cellcolor[HTML]{F2F7FD}70.21 & \cellcolor[HTML]{FFE0E0}43.62 \\
    Java $\rightarrow$ Py.
      & \cellcolor[HTML]{F8FBFE}55.32 & \cellcolor[HTML]{FFF2F2}24.47
      & \cellcolor[HTML]{E8F0FA}72.34 & \cellcolor[HTML]{FFE0E0}41.49
      & \cellcolor[HTML]{F2F7FD}69.15 & \cellcolor[HTML]{FFECEC}30.85
      & \cellcolor[HTML]{F8FBFE}60.64 & \cellcolor[HTML]{FFECEC}32.98
      & \cellcolor[HTML]{F8FBFE}62.77 & \cellcolor[HTML]{FFECEC}35.11 \\
    C++ $\rightarrow$ Java
      & \cellcolor[HTML]{E0ECF6}80.85 & \cellcolor[HTML]{FFFEFE}11.70
      & \cellcolor[HTML]{D4E6F1}94.19 & \cellcolor[HTML]{FFFEFE}15.96
      & \cellcolor[HTML]{D4E6F1}93.62 & \cellcolor[HTML]{FFFEFE}18.09
      & \cellcolor[HTML]{D4E6F1}92.47 & \cellcolor[HTML]{FFFEFE}13.98
      & \cellcolor[HTML]{E0ECF6}86.17 & \cellcolor[HTML]{FFFEFE}11.70 \\
    Py. $\rightarrow$ Java
      & \cellcolor[HTML]{E8F0FA}73.40 & \cellcolor[HTML]{FFFEFE}8.51
      & \cellcolor[HTML]{BEDBEE}95.35 & \cellcolor[HTML]{FFF2F2}20.21
      & \cellcolor[HTML]{D4E6F1}94.68 & \cellcolor[HTML]{FFF2F2}25.53
      & \cellcolor[HTML]{E0ECF6}81.91 & \cellcolor[HTML]{FFFEFE}18.09
      & \cellcolor[HTML]{D4E6F1}90.43 & \cellcolor[HTML]{FFF2F2}22.34 \\
    Java $\rightarrow$ C++
      & \cellcolor[HTML]{F8FBFE}60.64 & \cellcolor[HTML]{FFFEFE}3.19
      & \cellcolor[HTML]{BEDBEE}96.05 & \cellcolor[HTML]{FFFEFE}7.45
      & \cellcolor[HTML]{E0ECF6}88.30 & \cellcolor[HTML]{FFFEFE}12.77
      & \cellcolor[HTML]{F8FBFE}59.57 & \cellcolor[HTML]{FFFEFE}1.06
      & \cellcolor[HTML]{F8FBFE}58.51 & \cellcolor[HTML]{FFFEFE}1.06 \\
    Py. $\rightarrow$ C++
      & \cellcolor[HTML]{F8FBFE}45.74 & \cellcolor[HTML]{FFFEFE}1.06
      & \cellcolor[HTML]{BEDBEE}98.61 & \cellcolor[HTML]{FFFEFE}9.57
      & \cellcolor[HTML]{E0ECF6}87.23 & \cellcolor[HTML]{FFFEFE}7.45
      & \cellcolor[HTML]{F8FBFE}69.15 & \cellcolor[HTML]{FFFEFE}4.26
      & \cellcolor[HTML]{E0ECF6}80.85 & \cellcolor[HTML]{FFFEFE}2.13 \\
    \midrule
    Average
      & \cellcolor[HTML]{F8FBFE}60.46 & \cellcolor[HTML]{FFFEFE}12.06
      & \cellcolor[HTML]{E0ECF6}89.39 & \cellcolor[HTML]{FFF2F2}24.82
      & \cellcolor[HTML]{E0ECF6}83.51 & \cellcolor[HTML]{FFF2F2}22.87
      & \cellcolor[HTML]{E8F0FA}72.15 & \cellcolor[HTML]{FFFEFE}18.82
      & \cellcolor[HTML]{E8F0FA}74.82 & \cellcolor[HTML]{FFFEFE}19.33 \\
    \bottomrule
  \end{tabular}
  \end{threeparttable}
  }
\end{table}

\subsection{RQ4: Case Study of Translation Errors}
\label{sec:rq4}

Although existing methods achieve strong performance on the studied benchmarks, they still produce incorrect translations. To better understand these errors and identify the main challenges, we manually analyze the failed outputs generated by the studied methods. We adopt a two-step classification process. First, we randomly sample 30 failed cases to establish an initial set of error categories. Then, three evaluators, each with at least three years of software development experience, independently annotate the remaining cases in batches of 30. They discussed ambiguous cases and iteratively refined the coding categories. Under the final category criteria, the Fleiss'kappa~\cite{fleiss} coefficient was 0.9446, indicating substantial agreement among the three raters. Ultimately, all remaining disagreements were resolved through discussion. The final error taxonomy, shown in Figure~\ref{figs/type1}, divides errors into four major categories: syntax errors, static semantic errors, runtime errors, and logical errors. Each major category contains mutually exclusive subcategories with brief descriptions.

Each failed case is assigned to exactly one category. When a failure involves multiple error types, we classify it according to the following severity order: syntax errors have higher priority than static semantic errors, static semantic errors have higher priority than runtime errors, and runtime errors have higher priority than logical errors.

\begin{figure}[tbp]
\centering
\includegraphics[width=1.0\columnwidth]{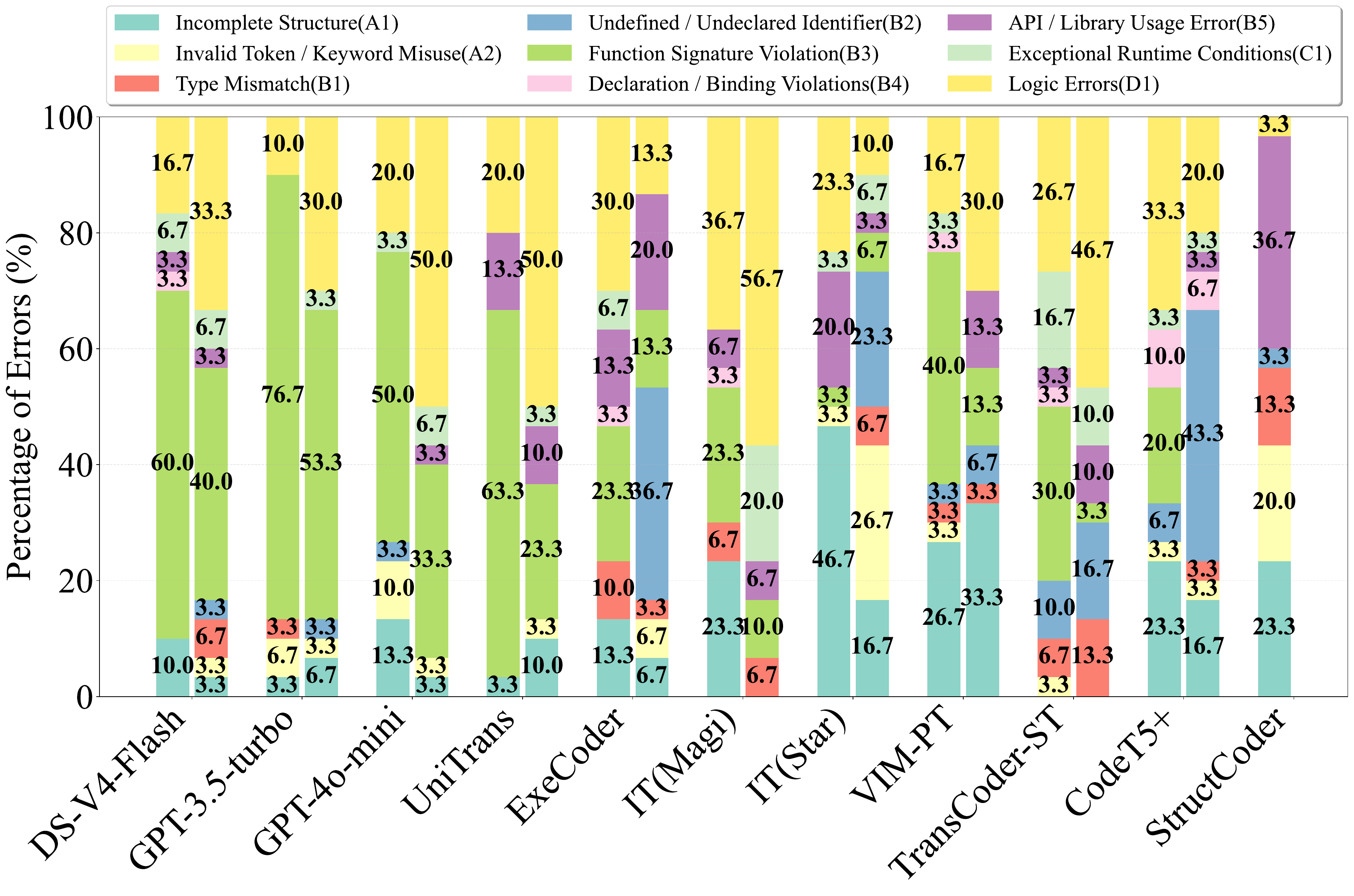}
\caption{Distribution of error types across datasets. The left side of each method represents the results on G-TransEval, and the right side represents the results on TransCoder-Uni.}
\label{figs/err}
\end{figure}

\begin{figure}[tbp]
\centering
\includegraphics[width=1.0\columnwidth]{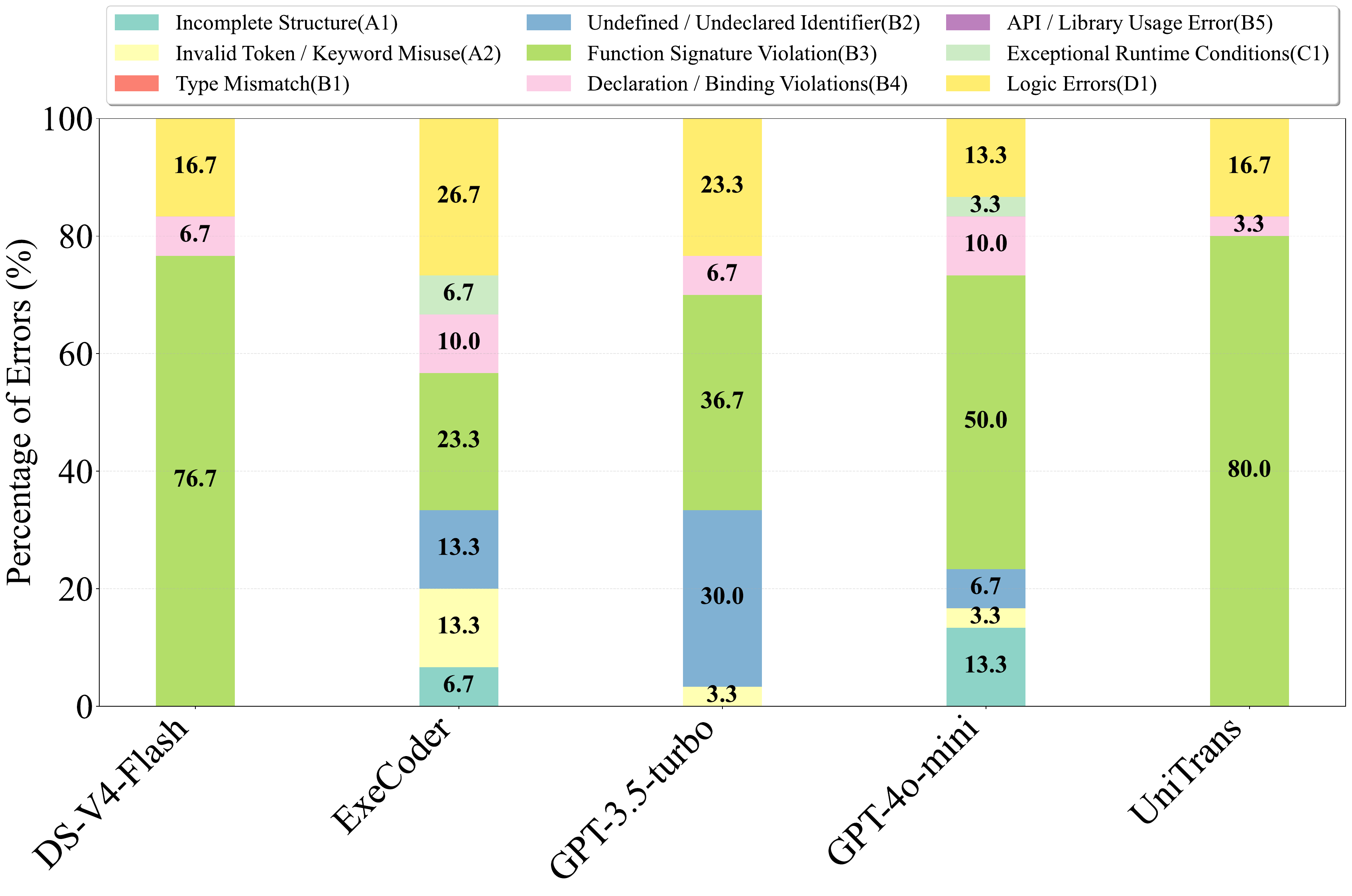}
\caption{Error case distribution on ClassEval-T.}
\label{figs/err_class}
\end{figure}

\subsubsection{RQ4.1 Method-Level Error Analysis}

For the method-level setting, we annotate 570 failed outputs generated by the studied methods. Specifically, we collect 30 failed cases for each method on each dataset, except that \StructCoder{} produces only 30 failed cases in total. The final distribution of error types is shown in Figure~\ref{figs/err}. Static semantic errors (49.37\%) and logical errors (27.46\%) account for the vast majority of all failed cases. This indicates that generating semantically and functionally correct programs remains the main challenge for code translation methods, including both learning-based and large language model (LLM)-based approaches. In contrast, syntax errors (18.25\%) and runtime errors (4.92\%) have a relatively smaller impact on the overall results. 

\begin{finding}
Among method-level failed cases, static semantic errors and logical errors are the most common, indicating that generating semantically and functionally correct programs remains a major challenge for existing code translation methods.
\end{finding}

Among static semantic errors, function signature violations are the most critical subcategory, accounting for 70.18\% of all static semantic errors. This problem arises because translation methods must implicitly infer type mappings between the source and target languages. It is particularly challenging when translating from an information-sparse source language to an information-rich target language. Even for state-of-the-art LLMs, accurately determining the appropriate target-language types requires a deep understanding of the semantics of the translated code.

\begin{figure}[tbp]
\centering
\includegraphics[width=1.0\columnwidth]{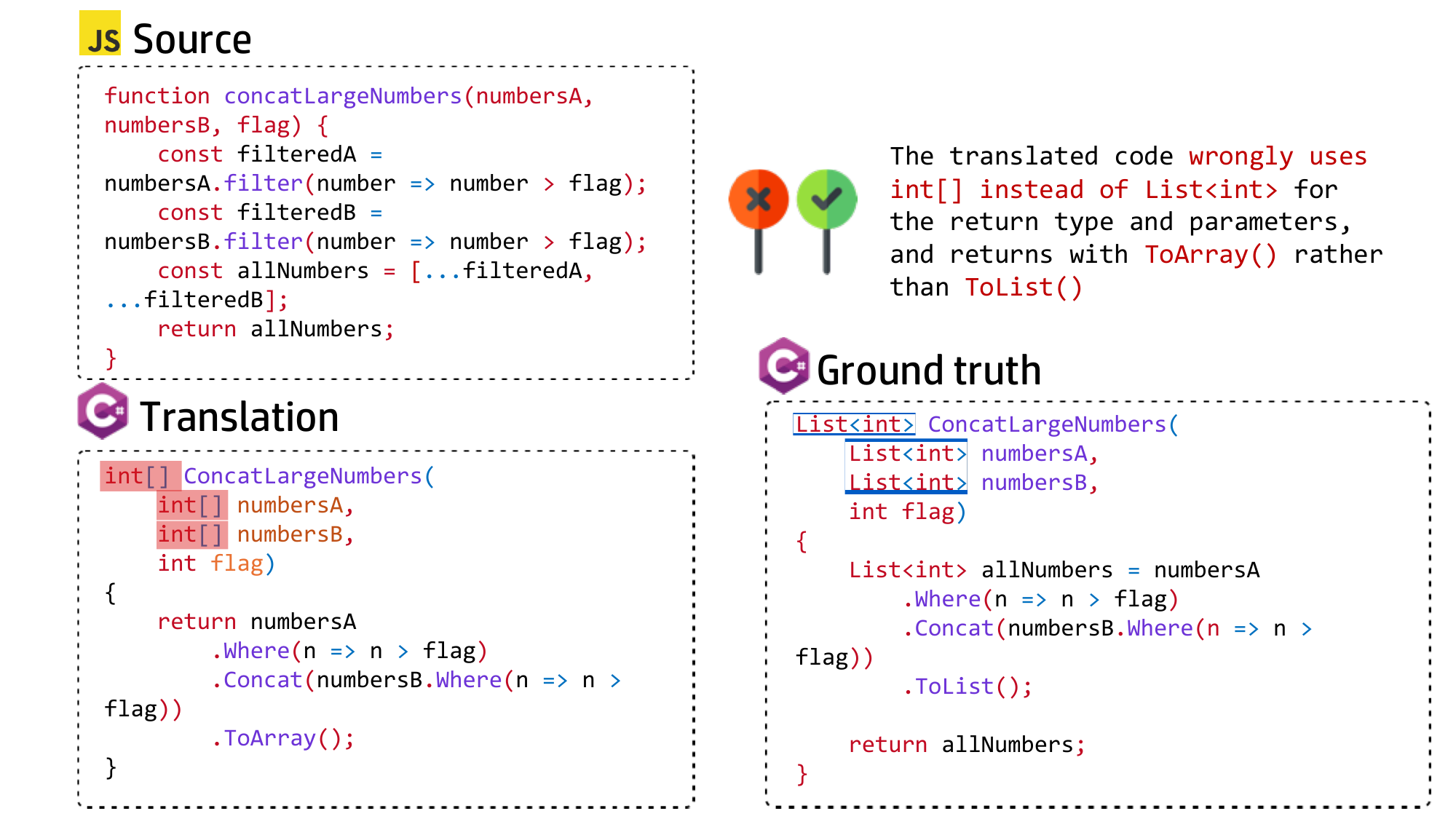}
\caption{An example of function signature violation}
\label{figs/exmp1}
\end{figure}

\begin{figure}[tbp]
\centering
\includegraphics[width=0.95\columnwidth]{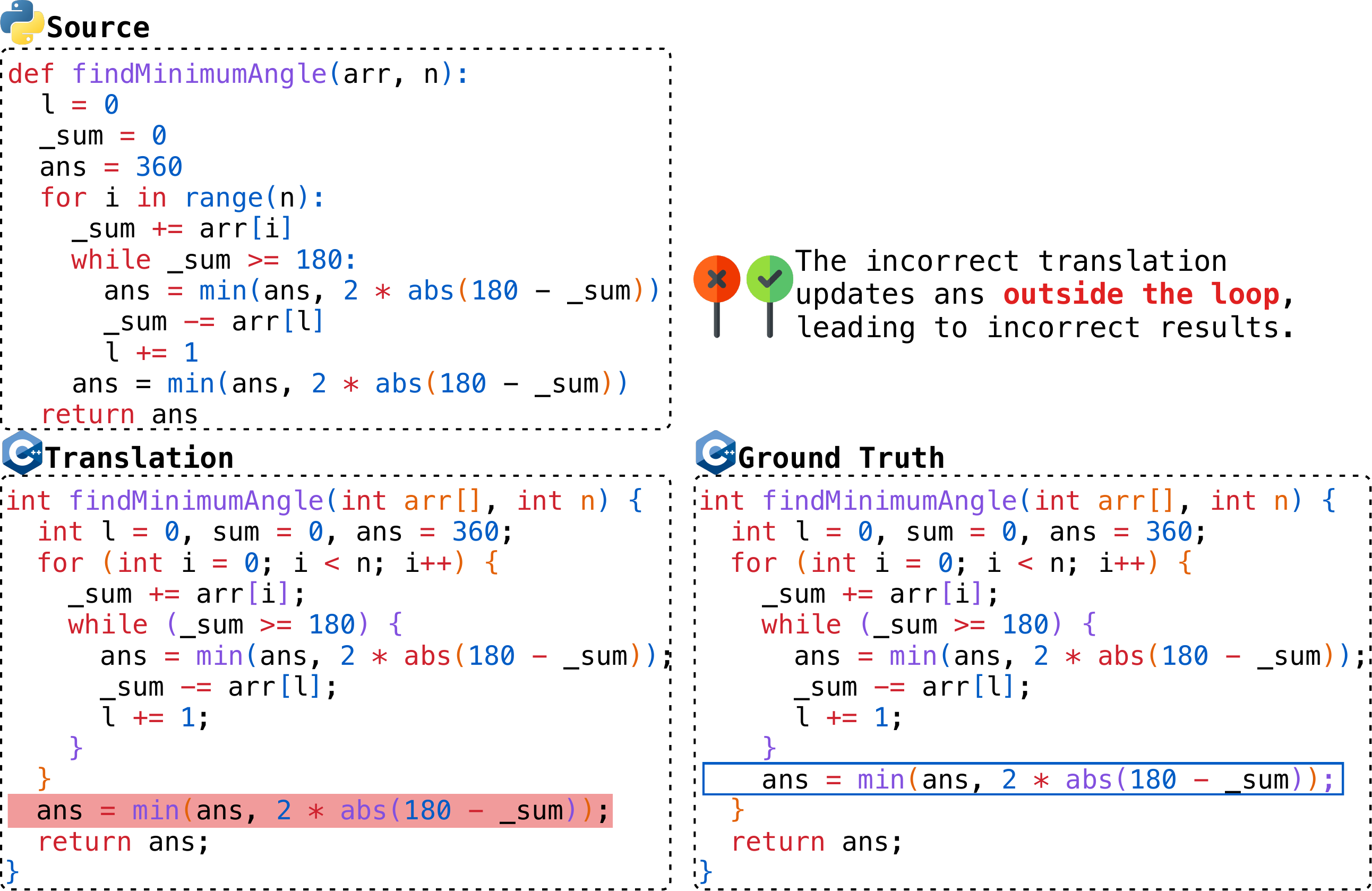}
\caption{An example of a logic error}
\label{figs/exmp2}
\end{figure}

For example, Figure~\ref{figs/exmp1} shows an incorrect translation from JavaScript to C\#, in which the generated method signature is incompatible with the expected interface. According to the reference answer, the return type and the first two parameter types of \texttt{ConcatLargeNumbers} should be \texttt{List<int>}. However, the translated code uses \texttt{int[]} for the return type and both list parameters, and correspondingly returns the result through \texttt{ToArray()} instead of \texttt{ToList()}. As a result, the generated method does not match the required interface and is incompatible with the benchmark tests. This example shows that, when translating from an information-sparse language such as JavaScript to a statically typed language such as C\#, accurately inferring collection types and mapping them to the expected target-language abstractions remains a key bottleneck.

For logical errors, we observe that many failures involving complex code logic are caused by misplaced statements. For example, Figure~\ref{figs/exmp2} shows a logical error made by \ITMagi{} when translating Python to C++. The assignment to the variable \texttt{ans} is incorrectly placed outside the loop, resulting in incorrect program behavior. Although the translated code can be compiled, it produces incorrect outputs. Such errors are difficult to detect, especially when the code involves complex algorithms, unless developers have a thorough understanding of the underlying logic.

Furthermore, we analyze the differences in error distributions between the two benchmarks. As shown in Figure~\ref{figs/err}, the distributions differ substantially, particularly for the two dominant error types discussed above: function signature violations and logical errors. Specifically, on G-TransEval, the average proportions of function signature violations and logical errors are 35.45\% and 21.52\%, respectively, whereas on TransCoder-Uni, the corresponding proportions are 19.67\% and 34.00\%.

To investigate the reasons behind these differences, we analyze the data types used in function signatures, which are the main source of function signature violations, as well as the corresponding code complexity, which is closely related to logical errors. The results are shown in Table~\ref{tab:type_comparison} and Table~\ref{tab:ccn}.

The results show that function signatures in G-TransEval have more complex type compositions and make extensive use of compound STL types. This increases the difficulty of cross-language translation because the translation method must not only infer the value types correctly but also understand the corresponding type-mapping rules. In contrast, the type structures in TransCoder-Uni are relatively simple and mainly consist of primitive types and C-style arrays, with primitive types alone accounting for 75.15\%. This difference in type complexity directly contributes to the substantially higher proportion of function signature violations in G-TransEval and also explains the cross-language translation difficulty observed in RQ2.

Similarly, the code complexity statistics show that TransCoder-Uni has a higher average cyclomatic complexity. Tasks with medium or higher difficulty account for nearly 30\% of the cases in TransCoder-Uni, compared with only 20.20\% in G-TransEval. This indicates that, although its type system is simpler, TransCoder-Uni contains more translation challenges involving complex logic, leading to more logical errors. This also explains why, on TransCoder-Uni, the performance gap between using Python, an information-sparse language, as the target language and using C++, an information-rich language, as the target language is smaller than that on G-TransEval. TransCoder-Uni contains more complex logic but simpler data types, which reduces the impact of inaccurate type translation.

\begin{finding}
As observed in Finding~3,
the performance difference between translating from statically to dynamically typed languages and translating in the reverse direction is closely related to the complexity of type inference and type mapping. More specifically, datasets containing more diverse and complex types are more susceptible to language effects.
\end{finding}

\subsubsection{RQ4.2 Class-Level Error Analysis}

To further examine whether the failure patterns identified above persist at a larger translation granularity, we extend the error analysis to class-level translation on ClassEval-T. In this setting, we annotate 150 failed outputs, with 30 cases generated by each studied method.

The distribution of class-level error cases is shown in Figure~\ref{figs/err_class}. The results show a more pronounced deviation in semantic correctness: static semantic errors account for 70.67 of all failed cases, while logical errors account for 19.33\%. In comparison, syntax errors and runtime errors account for only 8.00\% and 2.00\% of the cases, respectively. Thus, compared with method-level translation, class-level translation is less affected by syntactic problems but more constrained by semantic consistency, particularly the preservation of interfaces, member bindings, and cross-method dependencies.

Among static semantic errors, function signature violations remain the dominant subcategory, accounting for 75.47\% of all static semantic errors and 53.33\% of all class-level failed cases. The remaining static semantic errors mainly involve declaration or binding violations and undefined or undeclared identifiers. These results indicate that class-level translation is not merely a problem of mapping individual types. It also requires maintaining consistency across fields, methods, and class-level scopes. In other words, as the translation unit becomes larger, the model must preserve not only the correctness of local statements but also the structural relationships among the members of the class.

We further analyze the distribution of error types for each method in the class-level setting. For \UniTrans{} and the three commercial models, failures are primarily dominated by static semantic errors, most of which are function signature violations. In contrast, \ExeCoder{}, which uses a weaker underlying model, produces a more diverse range of error types, including more syntax errors and more frequent binding-related failures. This distribution suggests that stronger models are more capable of reducing syntax and runtime errors, while their remaining failures are more concentrated in static semantic and logical errors. A similar trend can also be observed in the method-level error analysis.

Overall, class-level and method-level code translation exhibit similar distributions of broad error categories, with static semantic errors and logical errors being the dominant types in both settings. However, class-level translation substantially amplifies the importance of type consistency and symbol binding. As the translation granularity increases, the main challenge shifts from local statement-level translation to maintaining syntactic and semantic consistency across the entire class. This challenge involves not only longer contexts and greater semantic complexity, but also stronger consistency requirements among class members, methods, and properties. These factors help explain why class-level translation becomes particularly difficult when the target language imposes stricter type and structural constraints.

\begin{finding}
Compared with method-level translation, class-level translation exhibits a similar overall error profile but substantially amplifies static semantic consistency problems. In particular, function signature violations account for 53.33\% of all class-level failed cases. This indicates that preserving interface consistency, symbol bindings, and cross-member dependencies becomes a primary bottleneck as the translation granularity increases.
\end{finding}

\begin{table}[tbp]
\centering
\caption{Statistical comparison of type distribution between the two datasets.}
\label{tab:type_comparison}
\begin{tabular}{lccccc}
\toprule
\multirow{2}{*}{Category} & \multicolumn{2}{c}{TransCoder-Uni} & \multicolumn{2}{c}{G-TransEval} \\
\cmidrule(lr){2-3} \cmidrule(lr){4-5}
                          & Count & \% of Total & Count & \% of Total \\
\midrule
Basic Types               & 1282  & 75.15\%     & 633   & 59.77\%     \\
STL Types                 & 122   & 7.15\%      & 426   & 40.23\%     \\
C-style Arrays            & 289   & 16.94\%     & 0     & 0\%         \\
Pointers/References       & 13    & 0.76\%      & 0     & 0\%         \\
\midrule
\textbf{Total}            & \textbf{1706} & \textbf{100\%} & \textbf{1059} & \textbf{100\%} \\
\bottomrule
\end{tabular}
\end{table}
\begin{table}[h]
  \centering
  \caption{Cyclomatic Complexity of the code for two datasets}
  \label{tab:ccn}
  
  \begin{tabular}{@{}l|c|c|c@{}}
    \toprule
    & \textbf{Metric} & \textbf{G-TransEval} & \textbf{TransCoder-Uni} \\
    \midrule
    
    \multirow{6}{*}{\rotatebox{90}{\textit{\makecell{Statistical\\Overview}}}} & Avg. Complexity & 4.14 & \textbf{4.57} \\
    & Median          & 4.00 & 4.00 \\
    & Std. Dev.       & 2.63 & 2.77 \\
    & Max / Min       & 19 / 1 & 19 / 1 \\
    & 90th Percentile & 7.00 & 8.00 \\
    \midrule
    
    \multirow{4}{*}{\rotatebox{90}{\textit{\makecell{Complexity\\Distri. (\%)}}}} & Simple (1--5)      & 79.8 & 70.3 \\
    & Medium (6--10)     & 17.2 & \textbf{26.9} \\
    & Complex (11--20)   & 3.0  & 2.8  \\
    & Very Complex ($>$20)& 0.0  & 0.0  \\
    \bottomrule
  \end{tabular}
\end{table}

\section{Implications}

Our comprehensive evaluation yields the following implications for future research in code translation.

\textbf{1) Future research should move beyond isolated method-level translation.}
In our experiments, LLMs and LLM-based methods generally outperform learning-based methods on correctness metrics in method-level translation. However, this advantage does not imply that method-level translation has been fully solved. More importantly, class-level translation remains substantially more difficult: the best-performing method achieves an average CA of only 24.82\%, despite obtaining an average CSR of 89.39\%. This large gap shows that generating compilable code is considerably easier than preserving semantic correctness in larger translation units. Future research should therefore investigate class-level, repository-level, and other practically relevant settings that require the preservation of inter-method dependencies, class structure, and global program semantics.

\textbf{2) Code translation should be evaluated using both correctness and similarity metrics.}
Our results show that learning-based methods may achieve CodeBLEU scores comparable to those of LLM-based methods, even though their CA and CSR are substantially lower. This indicates that similarity metrics alone cannot reliably reflect compilation success or functional correctness. At the same time, the smaller CodeBLEU gap between passed and failed outputs generated by LLM-based methods suggests that functionally correct translations may not always closely resemble reference implementations at the surface level. Future evaluations should therefore report correctness-oriented metrics together with similarity metrics and should avoid using CodeBLEU as a substitute for executable or test-based validation.

\textbf{3) Translation difficulty depends on both language direction and benchmark characteristics.}
Translation direction has a clear impact on performance: translating from information-rich languages (e.g., Java and C++) to information-sparse languages (e.g., Python and JavaScript) is generally easier than the reverse direction, which requires additional type inference and structural completion. Future methods should explicitly consider the challenges posed by different translation directions, particularly the type inference and type mapping issues involved when translating from information-sparse languages to information-rich languages, while also accounting for the type complexity and logical complexity of the target benchmark dataset.

\textbf{4) Static semantic consistency and logical equivalence remain the main bottlenecks.}
Among the sampled method-level failed cases, Static semantic errors (49.37\%) and logical errors (27.46\%) account for the vast majority of all failed cases. Function signature violations constitute 70.18\% of the static semantic errors and are especially frequent on G-TransEval, which contains more diverse and complex type compositions. At the class level, static semantic errors account for 70.67\% of the sampled failures, and function signature violations alone account for 53.33\% of all class-level failed cases. These findings highlight the need for stronger type-mapping constraints, interface checking, symbol-binding analysis, and semantic reasoning during translation. Future approaches should also improve their ability to preserve control-flow structure and algorithmic logic, since compilable code may still produce incorrect results.

\textbf{5) Class-level translation requires global consistency rather than local code generation.}
The class-level results show that increasing the translation granularity amplifies the influence of language-specific type systems, object models, declarations, and member dependencies. In particular, translations targeting C++ are considerably more difficult than those targeting Python or Java in several settings. This indicates that class-level translation cannot be addressed merely by extending the context window or translating methods independently. Future systems should maintain consistency across fields, method signatures, method bodies, inheritance relations, and cross-method references. Techniques such as structure-aware representations, dependency-aware translation, incremental verification, and class-level compilation or testing may therefore be important directions for improving practical code translation.

\begin{table}[tbp]
  \centering
  \caption{Data-leakage results and gaps against the original benchmarks.}
  \label{tab:data-leakage}
  \begin{threeparttable}
  \begin{tabular}{lcc}
    \toprule
    \textbf{Model} & \textbf{C++ $\rightarrow$ Python} & \textbf{C++ $\rightarrow$ Java} \\
    \midrule
    DS-V4-Flash
      & 94.00 / -5.25 / +4.70
      & 96.00 / -3.92 / +3.95 \\
    GPT-3.5-turbo
      & 94.00 / -2.17 / +7.58
      & 90.00 / -9.17 / -1.70 \\
    GPT-4o-mini
      & 92.00 / -2.75 / +6.58
      & 94.00 / -2.67 / +7.28 \\
    \bottomrule
  \end{tabular}
  \begin{tablenotes}
    \small
    \item Each cell reports \textit{Data-leakage Result / $\Delta$G / $\Delta$T}, where $\Delta$G and $\Delta$T denote gaps against G-TransEval and TransCoder-Uni, respectively. All values are percentages or percentage-point gaps.
  \end{tablenotes}
  \end{threeparttable}
\end{table}

\section{Threats to Validity}
\textbf{Internal Validity.} Our analysis involves expert judgment. Although three evaluators independently annotated the cases and reconciliation to reduce subjectivity (as mentioned in Section~\ref{sec:rq4}), the results may still be influenced by personal experience. Therefore, misinterpretations of translated code may introduce bias into the classification of error causes. To mitigate this threat, we used Fleiss' kappa~\cite{fleiss} to measure inter-rater agreement and assess the consistency of the classification results. We also adopted an inductive open-coding procedure~\cite{denzin2011sage}, in which the raters independently examined all error cases and resolved disagreements through discussion.

Another internal threat is data leakage. To mitigate this risk, for the learning-based methods under study, we manually cleaned the training samples that overlap with the test benchmarks. The open-source large language models used in this study (i.e., StarCoder2 and Magicoder) are primarily trained on data from GitHub repositories~\cite{starcoder, magicoder}, while our test benchmarks are sourced from the GeeksforGeeks platform, making direct overlap unlikely. This is also consistent with the discussions in prior works~\cite{unitrans, transcoder-st}. For the LLMs and LLM-based methods, we manually constructed a dataset containing 50 C++ code snippets along with their test suites, all selected from problems that appeared on Luogu~\cite{luogu} after the knowledge cutoff dates of the commercial LLMs. These problems appeared in the three commercial LLMs only after the cutoff date. On this dataset, we conducted experiments in both translation directions for the three LLMs we used, with results shown in the figure. Compared with the original results, we observed only limited increases or decreases. We consider this acceptable, as the data-leakage dataset differs from the original function-level dataset in terms of context length and problem difficulty. In summary, we believe that the impact of data leakage on the experiments in this paper is limited.


\textbf{External Validity.} Our findings may not fully generalize beyond the selected benchmarks, programming languages, and methods. Although we include diverse datasets and representative recent approaches, they may not cover all real-world translation scenarios, language pairs, or future models. To mitigate this threat, we evaluate multiple benchmarks with different properties and compare representative methods under a unified setting.

\section{Conclusion}

In this paper, we conducted a large-scale empirical study of 11 representative code translation techniques, including learning-based methods, LLM-based methods, and general-purpose LLMs. We evaluated them under unified settings on two method-level multilingual benchmarks and one class-level benchmark using both correctness and similarity metrics.

The results show that LLMs and LLM-based methods generally outperform learning-based methods in correctness, although their advantage is less evident on similarity metrics. Translation direction also affects performance, particularly on G-TransEval.
Furthermore, class-level translation remains substantially more difficult than method-level translation, especially in preserving semantic correctness and structural consistency. Our failure analysis shows that static semantic errors and logical errors are the dominant challenges, with function signature violations being particularly common.

Overall, our findings highlight the need for future code translation systems to improve \textit{type-aware}, \textit{structure-aware}, and \textit{context-aware} translation, while combining correctness-oriented and similarity-oriented evaluation.

\section{Data Availability}

The data and code supporting the findings of this study are available in our open-source repository: \artifact.

\bibliographystyle{IEEEtran}
\bibliography{ref}

\end{document}